\documentclass[12pt]{article}

\usepackage{amsmath}
\usepackage{cite}

\def\ti{\tilde}
\def\h{\hat}

\numberwithin{equation}{section}

\title{Supersymmetric higher spin models\\
in three dimensional spaces}

\author{I.L. Buchbinder${}^{ab}$\thanks{joseph@tspu.edu.ru},
T.V. Snegirev${}^{ac}$\thanks{snegirev@tspu.edu.ru}, Yu.M.
Zinoviev${}^{de}$\thanks{Yurii.Zinoviev@ihep.ru}
\\[0.5cm]
\it{\small ${}^a$Department of Theoretical Physics, Tomsk State
Pedagogical University,}\\
\it{\small Tomsk, 634061, Russia}\\[0.3cm]
\it{\small ${}^b$National Research Tomsk State University, Tomsk
634050, Russia}\\[0.3cm]
\it{\small ${}^c$National Research Tomsk Polytechnic University, Tomsk
634050, Russia}\\[0.3cm]
\it{\small ${}^d$Institute for High Energy Physics of National
Research Center "Kurchatov Institute"} \\
\it{\small Protvino, Moscow Region, 142281, Russia} \\[0.3cm]
\it\small{ ${}^e$Moscow Institute of Physics and Technology (State
University),} \\
\it{\small Dolgoprudny, Moscow Region, 141701, Russia}}

\date{}

\begin{document}

\maketitle

\begin{abstract}
We review the component Lagrangian construction of the
supersymmetric higher spin models in three dimensional (3D)
Minkowski and anti de Sitter ($AdS$) spaces. The approach is based on
the frame-like gauge invariant formulation, where massive higher
spin fields are realized through a system of massless ones. We
develop a supersymmetric generalization of this formulation to the
Lagrangian construction of the on-shell ${\cal N}=1$, 3D higher spin
supermultiplets. In 3D Minkowski space we show that the massive
supermultiplets can be constructed from one extended massless
supermultiplet by adding the mass terms to the Lagrangian and the
corresponding corrections to the supertransformations of the fermionic
fields. In 3D $AdS$ space we construct massive supermultiplets using a
formulation of the massive fields in terms of the set of gauge
invariant objects (curvatures) in the process of their consistent
supersymmetric deformation.
\end{abstract}

\thispagestyle{empty}
\newpage
\setcounter{page}{1}
\tableofcontents\pagebreak

\section{Introduction}

As soon as the supersymmetry (see e.g. the books \cite{1000},
\cite{BK}, \cite {GIOS}) has been discovered it was aroused an
immediate interest to finding the supersymmetric generalization of
the known theories. In short time there appeared the
supersymmetric extensions of such famous theories as the Standard
Model, the Einstein's gravity, string theory. Construction of
supersymmetric models and the study their properties on classical
and quantum levels became one of the most attractive trends in the
modern theoretical physics.

In the last decades there was an essential progress in the higher
spin field theory (see e.g. the reviews \cite{Vas04,BCIV05,BBS10}).
Purpose of this review is to describe a recent development of
Lagrangian construction for massless and massive supersymmetric
higher spin models in three dimensional Minkowski and anti de
Sitter spaces.

In the papers \cite{Cur79,Vas80} the massless ${\cal N}=1$
supersymmetric higher spin field theory was formulated in four
dimensional (4D) Minkowski space. The basic results of these papers
were the global on-shell ${\cal N}=1$ supertransformations leaving
invariant the pair of (Fang)-Fronsdal Lagrangians for free massless
higher spin-$(s,s+1/2)$ fields \cite{Frons78,FF78}. The off-shell
formulation of such system was given in \cite{KSP93,KS93} where
the ${\cal N}=1$ superfield extension of (Fang)-Fronsdal Lagrangians
in 4D Minkowski space was obtained. Later on this result was
generalized for 4D AdS space \cite{KS94}. In both cases the
constructed superfield models, up to eliminating of auxiliary fields,
reduce to the sum of spin-$s$ and spin-$(s+1/2)$ (Fang)-Fronsdal
Lagrangians thus describing ${\cal N}=1$,  4D massless higher spin
supermultiplets. Later making use the same technique the off-shell
formulation of 4D, ${\cal N}=2$ massless higher spin supermultiplets
was found \cite{GKS96a}.

There exist much less results in the study of supersymmetric massive
higher spin models. The reason is that shifting from massless
component formulation to the massive one we have to introduce the
very complicated higher derivative corrections to the
supertransformations. Moreover the higher the spin of the fields
entering supermultiplet the higher the number of derivatives one has
to consider. The problem of supersymmetric description of 4D massive
higher spin supermultiplet was resolved explicitly only in 2007 for
the case ${\cal N}=1$ on-shell Poincare superalgebra \cite{Zin07a}.
The solution was based on the generalization of the gauge invariant
formulation of the massive higher spin fields \cite{KZ97,Zin01,Met06}
to the case of massive supermultiplets. In such formulation the
massive supermultiplets are described as a system of the appropriate
massless ones coupled by local symmetries. On the other hand this
system of massless supermultiplets should be invariant under the
initial massless supertransformations corrected in a certain way. In
\cite{Zin07a} it was shown that to obtain the massive deformation it
is enough to add the non-derivative corrections to the
supertransformations for the fermions only. Complicated higher
derivative corrections to the supertransformations reappeared if one
tries to fix all local symmetries breaking gauge
invariance\footnote{The attempts to developed the off-shell superfield
formulation of the massive 4D higher spin supermultiplets were
considered for some examples in \cite{BG1}, \cite{BG2}, \cite{BG3}.}.
Surprisingly but in 4D the above results are still the main results in
massive supersymmetric higher spin theory till now.

Taking into account the difficulties in constructing the Lagrangian
formulation for 4D massive higher spin supermultiplets it is natural
to study the same problems in simpler case, for example to consider
the massive higher spin supermultiplets in three dimensions. An indeed
in the last years  much attention in the supersymmetric higher spin
theory has been focused on 3D spaces where higher spin theory is much
more simple (see e.g. \cite{Vas04,Blenc}). Here it is important to
emphasize that in general a supersymmetry in different dimensions is
realized quite differently. The matter is that the supersymmetry
operates with spinor fields which are formulated separately for each
space-time dimension. Therefore the 3D supersymmetry is independent
type of symmetry and should be considered by itself.

It it known that in 3D the massless higher spin fields ($k\geq3/2$)
do not propagate any physical degrees of freedom and one of the
reasons to study such models can be a possibility to consider their
deformation to a massive theory. In its turn the massive higher spin
fields in 3D do propagate two physical degrees of freedom
\cite{Bin82}.  It is important to note that massive higher spin fields
can be realized by different ways. One of the possibilities is to
generate the mass for 3D massless gauge fields by adding
Chern-Simons-like term \cite{PT89} generalizing 3D topologically
massive gravity \cite{DJT82}. Recently for these models the off-shell
${\cal N}=1$ and  ${\cal N}=2$ superfield extensions have been
constructed in 3D Minkowski space \cite{KT16,KO16}.

Another possibility to describe the 3D massive higher spin fields is
to add the usual mass terms being quadratic in physical fields without
derivatives. It mimics the corresponding construction for 4D
massive higher spin fields. This possibility has been realized in
the papers \cite{TV97,BSZ12a,BSZ14a}. The goal of this review is to
present the general methods of supersymmetric Lagrangian
construction for massive higher spin fields in 3D Minkowski and $AdS$
spaces. These methods are based on the gauge invariant description of
the massive fields \cite{BSZ12a,BSZ14a} and are realized in the
component approach for the case of on-shell ${\cal N}=1$
supersymmetry. Main content of this review is based on the papers
\cite{BSZ15,BSZ17}.

The review organized as follows. In the rest of the introduction we
fix our notations and conventions on 3D field variables. In sections 2
and 3 we present the Lagrangian formulation of 3D free bosonic and
fermionic higher spin fields respectively. In section 4 we construct
the Lagrangian formulation for higher spin supermultiplets in 3D
Minkowski space. Here we show that it is possible to construct one
extended massless supermultiplet and then smoothly deform it into the
massive one. Another approach is  used in section 5 for the
construction of the massive supermultiplets in 3D $AdS$. It is based
on the Lagrangian formulations in terms of the explicitly gauge
invariant objects and their consistent supersymmetric deformation.
Such approach is more elegant but it requires the introduction of the
so-called extra fields.

\noindent {\bf Notations and conventions.} In this review we use a
language of differential form when all the objects are some
$p$-forms $\Omega$ (p=0,1,2,3). It is defined as
$$
\Omega=\theta^{\mu_1}...\theta^{\mu_p}\Omega_{\mu_1...\mu_p},\qquad
\theta^\mu\theta^\nu=-\theta^\nu\theta^\mu
$$
In particular derivative is defined as 1-form
$d=\theta^\mu\partial_\mu$.

In 3D it is more convenient to use a frame-like multispinor
formalism where all the objects have totally symmetric local spinor
indices. To simplify the expressions we will use the condensed
notations for the spinor indices such that e.g.
$$
\Omega^{\alpha(2k)} = \Omega^{(\alpha_1\alpha_2 \dots \alpha_{2k})}
$$
Also we always assume that spinor indices denoted by the same
letters and placed on the same level are symmetrized, e.g.
$$
\Omega^{\alpha(2k)} \zeta^\alpha = \Omega^{(\alpha_1\dots
\alpha_{2k}} \zeta^{\alpha_{2k+1})}
$$
In flat space usual derivative $d$ commute $d\wedge d=0$ while for
$AdS$ space we use following normalization of the covariant
derivative:
$$
D \wedge D \zeta^\alpha = - \lambda^2 E^\alpha{}_\beta \zeta^\beta
$$
Basis elements of $1,2,3$-form spaces are $e^{\alpha(2)}$,
$E{}^{\alpha(2)}$, $E$ respectively where the last two are
defined as double and triple wedge product of $e^{\alpha(2)}$:
$$
e^{\alpha\alpha} \wedge e^{\beta\beta} =
\varepsilon^{\alpha\beta}{E}{}^{\alpha\beta}, \qquad
E{}^{\alpha\alpha} \wedge e^{\beta\beta} = \varepsilon^{\alpha\beta}
\varepsilon^{\alpha\beta} E.
$$
Also we write some useful relations for these basis elements
$$
E{}^\alpha{}_\gamma \wedge e^{\gamma\beta} = 3
\varepsilon^{\alpha\beta} E, \qquad e^\alpha{}_\gamma \wedge
e^{\gamma\beta} = 4 E{}^{\alpha\beta}.
$$
Further on the sign of wedge product $\wedge$ will be omitted.

\section{Free higher spin bosonic models}

In this section we review the Lagrangian description of the arbitrary
spin bosonic fields in three dimensional Minkowski and its
cosmological extension AdS spaces \cite{BSZ12a}. Both for massless and
massive fields we present the gauge invariant formulation using the
frame-like field variables. These fields generalize the tetrad and
Lorentz connection in the frame formulation of gravity. Such approach
allows us to construct gauge invariant objects (we will call them
curvatures) similar to the gravitational curvature and torsion and use
them to simplify many constructions.

\subsection{Massless fields}\label{massless_bosons}

It is well known that all massless fields with spin $k\geq1$ are
gauge ones therefore the gauge invariant formulation for them is a
natural form of description. As the gravity, which can be described
in term of the metric $g_{\mu\nu}$ field or in terms of the frame
field $e_\mu{}^a$ and the Lorentz connection $\omega_\mu{}^{a,b}$, the
massless higher spins can be described in two ways: metric-like or
frame-like. In 4D Minkowski or $AdS$ spaces the metric-like approach
leads to the Fronsdal formulation of massless integer spin-$k$ in
terms of totally symmetric tensors $\varphi_{\mu_1\mu_2...\mu_k}$
subject to the double tracelessness condition
$\varphi^\sigma{}_\sigma{}^\rho{}_\rho{}_{\mu_5...\mu_k}$ which
becomes nontrivial for $k\geq4$. In the frame-like approach such field
is described by generalized frame and Lorentz connection
fields\footnote{Actually for 4D massless fields with spin $k>2$ one
should consider extra gauge fields
$\Omega_\mu{}^{a_1...a_{k-1},b_1...b_t}$ where $2\leq t\leq(k-1)$.
They do not enter the free Lagrangian but do play a crucial role in
the Vasiliev interacting theory. However in 3D these extra fields are
absent in the massless case.}
$$
f_\mu{}^{a_1...a_{k-1}},\qquad\Omega_\mu{}^{a_1...a_{k-1},b}
$$
here $\mu$ is curved world index and $a,b$ are flat tangent indices.
World and flat indices are related by the background Minkowski or
$AdS$ frame $e_\mu{}^a$. The flat indices of these generalized fields
correspond to the irreducible $so(3,1)$ Lorentz tensors. The Fronsdal
formulation is recovered by eliminating the auxiliary field
$\Omega_\mu{}^{a_1...a_{k-1},b}$ and considering symmetric combination
$$
\varphi_{\mu_1...\mu_k}=e_{(\mu_1}{}^{a_1}e_{\mu_2}{}^{a_2}...e_{\mu_{k-1}}{}^{a_{k-1}}
f_{\mu_k)}{}_{a_1a_2...a_{k-1}}
$$

In 3D it is convenient to use a dual higher spin connection
$$
\Omega_\mu{}^{a_1...a_{k-1}}=\varepsilon_{bc}{}^{(a_1}
\Omega_\mu{}^{a_2...a_{k-1})b,c}
$$
where $\varepsilon^{abc}$ is a totally antisymmetric tensor. So in 3D
the frame-like and Lorentz-like higher spin gauge fields have the
same index structure and their local indices form irreducible
$so(2,1)$ Lorentz tensors. Due to the isomorphism $so(2,1)\sim sp(2)$
it is very convenient to use multispinor formalism in which our
gauge fields take form
$$
f_\mu{}^{\alpha_1\alpha_2...\alpha_{2k-2}},\qquad
\Omega_\mu{}^{\alpha_1\alpha_2...\alpha_{2k-2}}
$$
where $\alpha=1,2$ are spinor indices. Below we make use of language
of differential forms, considering the above higher spin field
variables as 1-forms (we omit index $\mu$), and condensed notations
for the spinor indices given in introduction.
\\
{\bf Spin-$(k+1)$} ($k\geq1$)
\\
In the frame-like formalism it is described by the physical 1-form
$f^{\alpha(2k)}$ and the auxiliary 1-form $\Omega^{\alpha(2k)}$.
Lagrangian in 3D $AdS$ looks like
\begin{eqnarray}\label{Lag_s_ml}
{\cal L} &=&  (-1)^{k+1} [ k \Omega_{\alpha(2k-1)\beta}
e^\beta{}_\gamma \Omega^{\alpha(2k-1)\gamma} + \Omega_{\alpha(2k)} D
f^{\alpha(2k)} \nonumber \\
&&\qquad\qquad+ \frac{k\lambda^2}{4} f_{\alpha(2k-1)\beta}
e^\beta{}_\gamma f^{\alpha(2k-1)\gamma}]
\end{eqnarray}
It is invariant under gauge transformations
\begin{eqnarray}
\delta \Omega^{\alpha(2k)} &=& D \eta^{\alpha(2k)}
 + \frac{\lambda^2}{4} e^\alpha{}_\beta \xi^{\alpha(2k-1)\beta}
\nonumber \\
\delta f^{\alpha(2k)} &=& D \xi^{\alpha(2k)} + e^\alpha{}_\beta
\eta^{\alpha(2k-1)\beta}
\end{eqnarray}
One can construct a pair of the gauge invariant curvatures
\begin{eqnarray}\label{Curv_s1}
{\cal{R}}^{\alpha(2k)} &=& D \Omega^{\alpha(2k)}
 + \frac{\lambda^2}{4} e^\alpha{}_\beta f^{\alpha(2k-1)\beta}
\nonumber \\
{\cal{T}}^{\alpha(2k)} &=& D f^{\alpha(2k)} + e^\alpha{}_\beta
\Omega^{\alpha(2k-1)\beta}
\end{eqnarray}
Using these curvatures the Lagrangian can be rewritten as follows
\begin{eqnarray}\label{Lag_s}
{\cal L} &=&  \frac{(-1)^{k+1}}{2}
[\Omega_{\alpha(2k)}{\cal{T}}^{\alpha(2k)}+f_{\alpha(2k)}{\cal{R}}^{\alpha(2k)}]
\end{eqnarray}
In order to obtain formulation in 3D Minkowski space one should put
$\lambda\rightarrow0$. Thus kinetic terms for higher spin fields is
just the first line in (\ref{Lag_s_ml}) where $D\rightarrow d$. Let us
also present kinetic terms for the lower spin fields. Frame-like
formulation for them is just the first-order formalism.
\\
{\bf Spin-1} is described by the physical 1-form $A$ and the auxiliary
0-form $B^{\alpha(2)}$. The Lagrangian has form
\begin{eqnarray*}
{\cal L} &=& E B_{\alpha\beta} B^{\alpha\beta} - B_{\alpha\beta}
e^{\alpha\beta} d A
\end{eqnarray*}
and it is invariant under the gauge transformations with 0-form
parameter $\xi$
$$
\delta A = d \xi
$$
\\
{\bf Spin-0} is described by the physical 0-form $\varphi$ and the
auxiliary 0-form $\pi^{\alpha(2)}$. The expression for the Lagrangian
looks like
\begin{eqnarray*}
{\cal L} &=& - E \pi_{\alpha\beta} \pi^{\alpha\beta} +
\pi_{\alpha\beta} E^{\alpha\beta} d \varphi
\end{eqnarray*}

As it will be seen below all these constructions play a role in the
gauge invariant formulation of the massive bosonic fields.

\subsection{Massive fields}\label{massive_boson}

In the gauge invariant form the massive spin $s$ field can be
described as a system of the massless fields with spins
$s,(s-1),...,0$. In the frame-like approach the corresponding set of
fields consists of
$$
(\Omega^{\alpha(2k)},f^{\alpha(2k)})\quad1\leq k\leq (s-1),\quad
(B^{\alpha(2)},A),\quad (\pi^{\alpha(2)},\varphi)
$$
The Lagrangian for the free fields with mass $m$ in 3D $AdS$ has the
form
\begin{eqnarray}\label{Lag_s}
{\cal L} &=& \sum_{k=1}^{s-1} (-1)^{k+1} [ k
\Omega_{\alpha(2k-1)\beta} e^\beta{}_\gamma
\Omega^{\alpha(2k-1)\gamma} + \Omega_{\alpha(2k)} D f^{\alpha(2k)}
 ] \nonumber \\
 && + E B_{\alpha\beta} B^{\alpha\beta} - B_{\alpha\beta}
e^{\alpha\beta} D A - E \pi_{\alpha\beta} \pi^{\alpha\beta} +
\pi_{\alpha\beta} E^{\alpha\beta} D \varphi \nonumber \\
 && + \sum_{k=1}^{s-2} (-1)^{k+1} a_k [ - \frac{(k+2)}{k}
\Omega_{\alpha(2)\beta(2k)} e^{\alpha(2)} f^{\beta(2k)} +
\Omega_{\alpha(2k)} e_{\beta(2)} f^{\alpha(2k)\beta(2)} ] \nonumber
\\
 && + 2a_0 \Omega_{\alpha(2)} e^{\alpha(2)} A - a_0
f_{\alpha\beta} E^\beta{}_\gamma B^{\alpha\gamma} + 2sM
\pi_{\alpha\beta} E^{\alpha\beta} A \nonumber \\
 && + \sum_{k=1}^{s-1} (-1)^{k+1} b_k f_{\alpha(2k-1)\beta}
e^\beta{}_\gamma f^{\alpha(2k-1)\gamma} + b_0 f_{\alpha(2)}
E^{\alpha(2)} \varphi + \frac{3a_0{}^2}{2}
 E \varphi^2
\end{eqnarray}
where
\begin{eqnarray}\label{boson_data}
a_k{}^2 &=& \dfrac{k(s+k+1)(s-k-1)}{2(k+1)(k+2)(2k+3)} [ M^2-(k+1)^2
\lambda^2 ] \nonumber \\
a_0{}^2 &=& \dfrac{(s+1)(s-1)}{3} [ M^2- \lambda^2 ] \\
b_k &=& \dfrac{s^2M^2}{4k(k+1)^2},\qquad b_0=\dfrac{sMa_0}{2},
\qquad M^2 = m^2 + (s-1)^2 \lambda^2\nonumber
\end{eqnarray}
Let us briefly discuss the structure of the Lagrangian (\ref{Lag_s}).
The first two lines are kinetic terms. It is just the sum of the
massless Lagrangians for spins $s,(s-1),...,0$ where ordinary
derivatives are replaced by the $AdS$ covariant ones. The third and
the fourth lines contain cross-terms for neighboring spins. These
cross-terms couple the individual massless fields into the whole
system describing the massive spin-$s$ field. Last line in
(\ref{Lag_s}) contains the mass terms. The coefficients in
(\ref{boson_data}) are determined by the invariance of the Lagrangian
under the following gauge transformations
\begin{eqnarray}\label{GT_ms}
\delta \Omega^{\alpha(2k)} &=& D \eta^{\alpha(2k)} +
\frac{(k+2)a_k}{k} e_{\beta(2)} \eta^{\alpha(2k)\beta(2)} \nonumber \\
 && + \frac{a_{k-1}}{k(2k-1)} e^{\alpha(2)} \eta^{\alpha(2k-2)}
 + \frac{b_k}{k} e^\alpha{}_\beta \xi^{\alpha(2k-1)\beta} \nonumber \\
\delta f^{\alpha(2k)} &=& D \xi^{\alpha(2k)} + e^\alpha{}_\beta
\eta^{\alpha(2k-1)\beta} + a_k e_{\beta(2)}
\xi^{\alpha(2k)\beta(2)} \nonumber \\
 && + \frac{(k+1)a_{k-1}}{k(k-1)(2k-1)} e^{\alpha(2)}
\xi^{\alpha(2k-2)} \nonumber \\
\delta \Omega^{\alpha(2)} &=& D \eta^{\alpha(2)} + 3a_1 e_{\beta(2)}
\eta^{\alpha(2)\beta(2)} + b_1
e^\alpha{}_\gamma \xi^{\alpha\gamma} \\
\delta f^{\alpha(2)} &=& D \xi^{\alpha(2)} + e^\alpha{}_\gamma
\eta^{\alpha\gamma} + a_1 e_{\beta(2)}
\xi^{\alpha(2)\beta(2)}  + 2a_0 e^{\alpha(2)} \xi \nonumber \\
\delta B^{\alpha(2)} &=& 2a_0 \eta^{\alpha(2)}, \qquad \delta A = D
\xi + \frac{a_0}{4} e_{\alpha(2)} \xi^{\alpha(2)}
\nonumber \\
\delta \pi^{\alpha(2)} &=& \frac{Msa_0}{2} \xi^{\alpha(2)}, \qquad
\delta \varphi = - 2Ms \xi \nonumber
\end{eqnarray}
This gauge invariant formulation of the massive theory in 3D $AdS$
space possesses some remarkable features. At first, we can consider a
flat limit $\lambda\rightarrow0$ and immediately obtain the
description of the massive fields in 3D Minkowski space. At second,
there is a correct massless limit $m\rightarrow0$ without the gap in
the number of physical degrees of freedom. In such limit our system
decomposes into two systems describing the massless spin-$s$ and the
massive spin-$(s-1)$ fields. At last, in $dS$ space when
$\lambda^2<0$ one can consider the so-called partially massless limits
$a_k\rightarrow0$. In such limit the system decomposes into the two
subsystems describing the partially massless spin-$s$ field and the
massive spin-$k$ field.

Now let us return to the general case of 3D $AdS$ massive spin-$s$
field. Having at our disposal the explicit expressions for the gauge
transformations (\ref{GT_ms}) we can construct the gauge invariant
curvatures. After the change of the normalization
\begin{eqnarray}\label{Norm_s}
B^{\alpha(2)}\rightarrow 2a_0B^{\alpha(2r)}\qquad
\pi^{\alpha(2)}\rightarrow b_0\pi^{\alpha(2)}
\end{eqnarray}
the first part of curvatures looks like
\begin{eqnarray}\label{Curv_s1}
{\cal{R}}^{\alpha(2k)} &=& D \Omega^{\alpha(2k)} +
\frac{(k+2)a_k}{k} e_{\beta(2)} \Omega^{\alpha(2k)\beta(2)}
\nonumber
\\
 && + \frac{a_{k-1}}{k(2k-1)} e^{\alpha(2)} \Omega^{\alpha(2k-2)}
 + \frac{b_k}{k} e^\alpha{}_\beta f^{\alpha(2k-1)\beta} \nonumber \\
{\cal{T}}^{\alpha(2k)} &=& D f^{\alpha(2k)} + e^\alpha{}_\beta
\Omega^{\alpha(2k-1)\beta} + a_k e_{\beta(2)}
f^{\alpha(2k)\beta(2)} \nonumber \\
 && + \frac{(k+1)a_{k-1}}{k(k-1)(2k-1)} e^{\alpha(2)}
f^{\alpha(2k-2)} \nonumber \\
{\cal{R}}^{\alpha(2)} &=& D \Omega^{\alpha(2)} + 3a_1 e_{\beta(2)}
\Omega^{\alpha(2)\beta(2)} + b_1 e^\alpha{}_\gamma
f^{\alpha\gamma}-{a_0}^2E^\alpha{}_\beta B^{\alpha\beta}
+b_0E^{\alpha(2)}\varphi \\
{\cal{T}}^{\alpha(2)} &=& Df^{\alpha(2)} + e^\alpha{}_\gamma
\Omega^{\alpha\gamma} + a_1 e_{\beta(2)}
f^{\alpha(2)\beta(2)}  + 2a_0 e^{\alpha(2)}A \nonumber \\
{\cal{A}} &=& DA + \frac{a_0}{4} e_{\alpha(2)}
f^{\alpha(2)}-2a_0E_{\gamma(2)}B^{\gamma(2)}
\nonumber\\
\Phi &=&D\varphi+ 2MsA-b_0e_{\alpha(2)}\pi^{\alpha(2)} \nonumber
\end{eqnarray}
There is a peculiarity when we try to construct the curvatures for the
$B^{\alpha(2)} and \pi^{\alpha(2)}$ fields. Namely, in order to
achieve gauge invariance for them we should introduce the so-called
extra fields $B^{\alpha(4)},\pi^{\alpha(4)}$ with the following gauge
transformations:
$$
\delta B^{\alpha(4)}=\eta^{\alpha(4)}\qquad \delta
\pi^{\alpha(4)}=\xi^{\alpha(4)}
$$
Then the corresponding gauge invariant curvatures look like
\begin{eqnarray}
{\cal{B}}^{\alpha(2)}
&=&DB^{\alpha(2)}-\Omega^{\alpha(2)}+{b_1}e^\alpha{}_\beta\pi^{\alpha\beta}+
3a_1e_{\beta(2)}B^{\alpha(2)\beta(2)}
\nonumber\\
\Pi^{\alpha(2)} &=&D\pi^{\alpha(2)}-f^{\alpha(2)}+e^\alpha{}_\beta
B^{\alpha\beta}-\frac{a_0}{sM}e^{\alpha(2)}\varphi+a_1e_{\beta(2)}\pi^{\alpha(2)\beta(2)}
\end{eqnarray}
In turn to construct gauge invariant curvatures for the
$B^{\alpha(4)},\pi^{\alpha(4)}$ we should introduce the extra fields
$B^{\alpha(6)},\pi^{\alpha(6)}$ and so on. The procedure ends when
we construct curvatures for $B^{\alpha(2s-2)},\pi^{\alpha(2s-2)}$.
Thus the full set of extra fields is
$B^{\alpha(2k)},\pi^{\alpha(2k)}$, $2\leq k\leq s-1$ with the
following gauge transformations
$$
\delta B^{\alpha(2k)}=\eta^{\alpha(2k)}\qquad \delta
\pi^{\alpha(2k)}=\xi^{\alpha(2k)}
$$
and the gauge invariant curvatures
\begin{eqnarray}\label{Curv_s2}
{\cal{B}}^{\alpha(2k)}
&=&DB^{\alpha(2k)}-\Omega^{\alpha(2k)}+\frac{b_k}{k}e^\alpha{}_\beta\pi^{\alpha(2k-1)\beta}+
\frac{a_{k-1}}{k(2k-1)}e^{\alpha(2)}B^{\alpha(2k-2)}\nonumber\\
&&+\frac{(k+2)}{k}a_ke_{\beta(2)}B^{\alpha(2k)\beta(2)} \nonumber
\nonumber\\
\Pi^{\alpha(2k)}
&=&D\pi^{\alpha(2k)}-f^{\alpha(2k)}+e^\alpha{}_\beta
B^{\alpha(2k-1)\beta}+\frac{(k+1)a_{k-1}}{k(k-1)(2k-1)}e^{\alpha(2)}\pi^{\alpha(2k-2)}\nonumber\\
&&+a_ke_{\beta(2)}\pi^{\alpha(2k)\beta(2)}
\end{eqnarray}

In three dimensions it is possible to rewrite the Lagrangian in terms
of the curvatures only \cite{Zin16}. In the case of arbitrary integer
spin field, the corresponding Lagrangian (\ref{Lag_s}) can be
rewritten in the following simple form
\begin{eqnarray}\label{LagC_s}
{\cal{L}}&=&-\frac{1}{2}\sum_{k=1}^{s-1}(-1)^{k+1}[{\cal{R}}_{\alpha(2k)}\Pi^{\alpha(2k)}
+{\cal{T}}_{\alpha(2k)}{\cal{B}}^{\alpha(2k)}]
+\frac{a_0}{2sM}e_{\alpha(2)}{\cal{B}}^{\alpha(2)}\Phi
\end{eqnarray}

Thus there are two approaches to the construction of the
supersymmetric (or interacting in general case) higher spin models.
According to one of them one can work with the explicit field
variables and the Lagrangian in the form (\ref{Lag_s}). It is
straightforward but rather cumbersome way. We will use it in section 4
for the more simple case of 3D Minkowski space. According to the other
way one can work in terms of the gauge invariant curvatures and
Lagrangian in form (\ref{LagC_s}). It is more elegant way and we use
it in section 5 to study the supersymmetric higher spin models in 3D
$AdS$.

\section{Free higher spin fermionic models}

In this section we review the Lagrangian description of arbitrary spin
fermionic fields in three dimensional Minkowski and its cosmological
extension $AdS$ spaces \cite{BSZ14a}. As in the bosonic case we
present frame-like gauge invariant formulation. It is natural form
of description for massless fields and for massive fields gauge
invariant formulation is realized as a system of massless fields
coupled by Stueckelberg symmetries.

\subsection{Massless fields}\label{massless_fermions}

As in the integer spin case all massless fields with half-integer spin
$(k+1/2)\geq3/2$ are gauge ones and can be described according to the
metric-like or the frame-like approaches. In 4D Minkowski or $AdS$
spaces the metric-like approach leads to Fang-Fronsdal formulation of
the massless half-integer spin-$(k+1/2)$ fields in terms of totally
symmetric spin-tensors $\psi_{\mu_1\mu_2...\mu_k,\alpha}$ (here
$\alpha,\beta=1,2$ is spinor index) subject to the
$\gamma$-tracelessness condition
$\gamma^\sigma{}_\alpha{}^\beta\psi^\rho{}_{\rho\sigma}{}_{\mu_3...\mu_k,\beta}$
which becomes nontrivial for $k\geq3$. In the frame-like approach such
field is described by generalized frame-like field\footnote{As in the
bosonic case in 4D for the massless fields with half-integer spins
$(k+1/2)>3/2$ one should consider extra gauge fields
$\Phi_\mu{}^{a_1...a_{k-1},b_1...b_t,\alpha}$ where $2\leq
t\leq(k-1)$. They play a crucial role in Vasiliev interacting
theory. In 3D these extra fields vanishes in the massless case.}
$$
\Phi_\mu{}^{a_1...a_{k-1},\alpha}
$$
here $\mu$ is curved world index and $a_i$ are Lorentz flat  tangent
indices. World and flat indices are related by background Minkowski or
$AdS$ frame $e_\mu{}^a$. Their flat indices correspond to the
irreducible $so(3,1)$ Lorentz spin-tensors. The Fang-Fronsdal
spin-tensors are recovered by considering symmetric combination
$$
\psi_{\mu_1...\mu_k,\alpha}=e_{(\mu_1}{}^{a_1}e_{\mu_2}{}^{a_2}...e_{\mu_{k-1}}{}^{a_{k-1}}
\Phi_{\mu_k)}{}_{a_1a_2...a_{k-1},\alpha}
$$

Again due to the isomorphism $so(2,1)\sim sp(2)$ it is more
convenient to use multispinor formalism in which our gauge field
take the form
$$
\Phi_\mu{}^{\alpha_1\alpha_2...\alpha_{2k-1}}
$$
Below we make use of the language of differential forms, considering
the higher spin field variables as 1-forms, and using condensed
notations for the spinor indices given in introduction.
\\
{\bf Spin $k+3/2$} ($k\geq0$)
\\
In the frame-like formalism it is described by physical 1-form
$\Phi^{\alpha(2k-1)}$. The Lagrangian in 3D $AdS$ has the following
form
\begin{eqnarray}\label{Lag_s/2}
{\cal L} &=&  i\frac{(-1)^{k+1}}{2} [\Phi_{\alpha(2k+1)} D
\Phi^{\alpha(2k+1)}  + \frac{(2k+1)\lambda}{2}
\Phi_{\alpha(2k)\beta} e^\beta{}_\gamma \Phi^{\alpha(2k)\gamma}]
\end{eqnarray}
It is invariant under the following gauge transformations
\begin{eqnarray}\label{GT_s/2}
\delta \Phi^{\alpha(2k+1)} &=& D \xi^{\alpha(2k+1)} +
\frac{\lambda}{2} e^\alpha{}_\beta \xi^{\alpha(2k)\beta}
\end{eqnarray}
The gauge invariant curvature looks like
\begin{eqnarray}\label{Curv_s/2}
{\cal{F}}^{\alpha(2k+1)} &=& D \Phi^{\alpha(2k+1)} +
\frac{\lambda}{2} e^\alpha{}_\beta \Phi^{\alpha(2k)\beta}
\end{eqnarray}
Using this curvature the Lagrangian can be rewritten as follows
\begin{eqnarray}
{\cal L} &=&  i\frac{(-1)^{k+1}}{2}
\Phi_{\alpha(2k+1)}{\cal{F}}^{\alpha(2k+1)}
\end{eqnarray}
3D Minkowski case corresponds to the flat limit $\lambda\rightarrow0$.
Let us also write out kinetic term for the spin-$1/2$ field.
\\
{\bf Spin 1/2} is described by the physical 0-form $\phi^\alpha$. It
is not a gauge field and the Lagrangian looks like
\begin{eqnarray*}
{\cal L} &=& \frac{1}{2} \phi_\alpha E^\alpha{}_\beta d \phi^\beta
\end{eqnarray*}

\subsection{Massive fields}\label{massive_fermion}

To describe the massive spin-$(s+1/2)$ field in the gauge invariant
form we have to consider a system of the massless fields with spins
$(s+1/2),(s-1/2),...,1/2$. In the frame-like approach the
corresponding set of fields consists of
$$
\Phi^{\alpha(2k+1)}\quad0\leq k\leq (s-1),\quad \phi^\alpha
$$
Lagrangian for the free field with mass $m_1$ in 3D AdS space looks
like
\begin{eqnarray}\label{Lag_s/2}
\frac{1}{i} {\cal L} &=& \sum_{k=0}^{s-1} (-1)^{k+1} [ \frac{1}{2}
\Phi_{\alpha(2k+1)} D \Phi^{\alpha(2k+1)} ]
+ \frac{1}{2} \phi_\alpha E^\alpha{}_\beta D \phi^\beta \nonumber \\
 && + \sum_{k=1}^{s-1} (-1)^{k+1} c_k \Phi_{\alpha(2k-1)\beta(2)}
e^{\beta(2)} \Phi^{\alpha(2k-1)} + c_0 \Phi_\alpha E^\alpha{}_\beta
\phi^\beta \nonumber \\
 && + \sum_{k=0}^{s-1} (-1)^{k+1} \frac{d_k}{2}
\Phi_{\alpha(2k)\beta} e^\beta{}_\gamma \Phi^{\alpha(2k)\gamma} -
\frac{3d_0}{2} E \phi_\alpha \phi^\alpha
\end{eqnarray}
where
\begin{eqnarray}\label{fermion_data}
c_k{}^2 &=& \dfrac{(s+k+1)(s-k)}{2(k+1)(2k+1)} [ M_1{}^2 - (2k+1)^2
\frac{\lambda^2}{4} ] \nonumber\\
c_0{}^2 &=& 2s(s+1) [ M_1{}^2 - \frac{\lambda^2}{4} ]
\\
d_k &=& \dfrac{(2s+1)}{(2k+3)} M_1, \qquad M_1{}^2 = m_1{}^2 +
(s-\frac{1}{2})^2 \lambda^2\nonumber
\end{eqnarray}
The structure of Lagrangian (\ref{Lag_s/2}) is the same as in the
bosonic case. The first line is kinetic terms, the second line is
cross-terms and the third line is mass terms. The coefficients in
(\ref{fermion_data}) are determined by requirement of the invariance
of the Lagrangian under the following gauge transformations
\begin{eqnarray}\label{GT_s/2}
\delta \Phi^{\alpha(2k+1)} &=& D \xi^{\alpha(2k+1)} +
\frac{d_k}{(2k+1)} e^\alpha{}_\beta \xi^{\alpha(2k)\beta} \nonumber \\
 && + \frac{c_k}{k(2k+1)} e^{\alpha(2)} \xi^{\alpha(2k-1)}
 + c_{k+1} e_{\beta(2)} \xi^{\alpha(2k+1)\beta(2)} \\
\delta \phi^\alpha &=& c_0 \xi^\alpha \nonumber
\end{eqnarray}
In such formulation we can take correct massless limit
$m_1\rightarrow0$ in $AdS$ ($\lambda^2>0$) and correct partially
massless limits $c_k\rightarrow0$ in $dS$ ($\lambda^2<0$). Taking flat
limit $\lambda\rightarrow0$ we obtain the description of the massive
field in 3D Minkowski space. Note that the Lagrangian (\ref{Lag_s/2})
describes massive Majorana left fermion carrying one physical degree
of freedom.

Let us return to the general massive fermion and reformulate the
theory in terms of the gauge invariant curvatures. Having at our
disposal the explicit expressions for the gauge transformations
(\ref{GT_s/2}) we can construct the gauge invariant objects. After
change of normalization
\begin{eqnarray}\label{NormF}
\phi^{\alpha}\rightarrow c_0\phi^{\alpha}
\end{eqnarray}
they take the form
\begin{eqnarray}\label{Curv_s/2}
{\cal{F}}^{\alpha(2k+1)} &=& D \Phi^{\alpha(2k+1)} +
\frac{d_k}{(2k+1)} e^\alpha{}_\beta \Phi^{\alpha(2k)\beta} \nonumber
\\
 && + \frac{c_k}{k(2k+1)} e^{\alpha(2)} \Phi^{\alpha(2k-1)}
 + c_{k+1} e_{\beta(2)} \Phi^{\alpha(2k+1)\beta(2)} \nonumber\\
{\cal{F}}^{\alpha} &=& D \Phi^{\alpha} + {d_0} e^\alpha{}_\beta
\Phi^{\beta} + c_{1} e_{\beta(2)} \Phi^{\alpha\beta(2)}
-c_0{}^2E^\alpha{}_\beta\phi^\beta
\nonumber\\
{\cal{C}}^\alpha&=&D\phi^\alpha-\Phi^\alpha+d_0e^\alpha{}_\beta\phi^\beta
+c_1e_{\beta(2)}\phi^{\alpha\beta(2)}
\end{eqnarray}
As in the case of integer spins in order to achieve gauge invariance
for ${\cal{C}}^\alpha$ we have introduced extra 0-form
$\phi^{\alpha(3)}$ with the gauge transformations
$$
\delta\phi^{\alpha(3)}=\xi^{\alpha(3)}
$$
In turn to construct gauge invariant curvatures for
$\phi^{\alpha(3)}$ field we should introduce extra 0-form
$\phi^{\alpha(5)}$ and so on. Iterations ends at the case of
$\phi^{\alpha(2s-1)}$ so that the full set of extra fields we should
introduce is $\phi^{\alpha(2k+1)}$, $1\leq k\leq (s-1)$ with the
following gauge transformations
$$
\delta\phi^{\alpha(2k+1)}=\xi^{\alpha(2k+1)}
$$
The gauge invariant curvatures for them have the form
\begin{eqnarray}
{\cal{C}}^{\alpha(2k+1)}&=&D\phi^{\alpha(2k+1)}-\Phi^{\alpha(2k+1)}
+\frac{d_k}{(2k+1)}e^\alpha{}_\beta\phi^{\alpha(2k)\beta}\nonumber\\
&& +\frac{c_k}{k(2k+1)}e^{\alpha(2)}\phi^{\alpha(2k-1)}
+c_{k+1}e_{\beta(2)}\phi^{\alpha(2k+1)\beta(2)}
\end{eqnarray}
Finally the Lagrangian (\ref{Lag_s/2}) can be rewritten in terms of
these curvatures as follows
\begin{eqnarray}\label{LagC_s/2}
{\cal{L}}&=&-\frac{i}{2}\sum_{k=0}^{s-2}(-1)^{k+1}{\cal{F}}_{\alpha(2k+1)}{\cal{C}}^{\alpha(2k+1)}
\end{eqnarray}

In the next sections we  study a supersymmetric higher spin
models. In 3D Minkowski space we use the gauge invariant
formulation in terms of the explicit fields and the Lagrangian in the
form (\ref{Lag_s/2}). In 3D $AdS$ space we use the formulation in
terms of the gauge invariant curvatures and the Lagrangian in the form
(\ref{LagC_s/2}).

\section{Lagrangian construction of higher spin supermultiplets in 3D
Minkowski space}

In this section we show how to combine the bosonic and the
fermionic higher spin fields into one supermultiplet in 3D Minkowski
space and restrict ourselves with ${\cal N}=1$ on-shell
supersymmetry. Our construction is based on the gauge invariant
formulation given above in terms of the field variables. As we have
shown in such formulation massive field in the massless limit
decomposes into system of the massless fields. If we take such
decomposition for each field in the massive supermultiplet we obtain
in general its decomposition into supersymmetric system of the
massless fields. So the main idea is to start with this supersymmetric
system of the massless fields and construct smooth massive
deformation. In other words we generalize the gauge invariant
formulation to the supersymmetric case.

\subsection{3D vs 4D supermultiplets}

For the first time the idea to construct the massive higher spin
supermultiplets from the supersymmetric system of massless fields
was realized for 4D Minkowski space \cite{Zin07a}. Moreover it was
shown that this supersymmetric system of the massless fields is
perfectly combined into the system of the massless supermultiplets.
Therefore it is useful to consider how the familiar massive 4D
supermultiplets decomposes into the massless ones and then compare it
with decomposition of the more specific massive 3D supermultiplets. In
both cases it is used that in the gauge invariant formulation massive
bosonic and fermionic fields decompose into a set of the massless ones
\begin{eqnarray}\label{FD}
s&\xrightarrow{m=0}& s\oplus(s-1)\oplus...\oplus0=\sum_{k=0}^{s}k
\nonumber\\
(s+\frac12)&\xrightarrow{m=0}&
(s+\frac12)\oplus(s-\frac12)\oplus...\oplus\frac12=\sum_{k=0}^{s}(k+\frac12)
\end{eqnarray}

Massive 4D, ${\cal N}=1$ supermultiplets with the half-integer
spin-$(s+1/2)$ as the highest one contains four massive fields $s+1/2,
s,s'$ and $s-1/2$. Recall that in 4D massive bosonic spin-$s$ has
$2s+1$ d.o.f. and massive fermionic spin-$(s+1/2)$ has $2s+2$ d.o.f.
All 4D massless fields have 2 d.o.f. except for the spin 0 which has
1. In
the massless limit massive supermultiplet decomposes into massless
ones in the same way as (\ref{FD}). Simple counting d.o.f. gives
\begin{eqnarray}
\left( \begin{array}{ccc}
 & {s+\frac{1}{2}} & \\ s & & s' \\  & {s-\frac{1}{2}} &
\end{array} \right) &\xrightarrow{m=0}&\sum_{k=1}^s \left(
\begin{array}{c} {k+\frac{1}{2}} \\ k \end{array}
\right)\oplus \sum_{k=1}^s \left( \begin{array}{c} {k'}
\\ k-\frac12 \end{array} \right)\oplus\left( \begin{array}{c}
{\frac{1}{2}}
\\ 0,0'
\end{array} \right)
\end{eqnarray}
So to construct massive supermultiplets we have to start with $2s+1$
massless ones and find a massive deformation. However in
\cite{Zin07a} it was shown that the crucial point in the whole
construction is the possibility to make a dual mixing of the massless
supermultiplets by rotating fields with spin $k$ and spin $k'$
\begin{eqnarray}\label{mix}
\left( \begin{array}{c} {k+\frac{1}{2}} \\
k \end{array} \right)\oplus  \left( \begin{array}{c} {k'}
\\ k-\frac12 \end{array} \right)\quad\rightarrow\quad
\left( \begin{array}{ccc}
 & {k+\frac{1}{2}} & \\ k & & k' \\  & {k-\frac{1}{2}} &
\end{array} \right)
\end{eqnarray}
In such mixing the massless bosonic fields with spin $k$ and $k'$
must have equal spins but opposite parities. Thus the structure of the
decomposition of the 4D massive supermultiplets with higher
half-integer spin into the massless ones looks like
\begin{eqnarray*}
\left( \begin{array}{ccc}
 & {s+\frac{1}{2}} & \\ s & & s' \\  & {s-\frac{1}{2}} &
\end{array} \right) &\xrightarrow{m=0}&\sum_{k=1}^s \left(
\begin{array}{ccc}
 & {k+\frac{1}{2}} & \\ k & & k' \\  & {k-\frac{1}{2}} &
\end{array} \right)\oplus\left( \begin{array}{c} {\frac{1}{2}} \\ 0,0'
\end{array}
\right)
\end{eqnarray*}
Analogously we obtain that the 4D massive supermultiplets with higher
integer spin in the massless limit has decomposition
\begin{eqnarray*}
\left( \begin{array}{ccc}
 & {s+1} & \\ s+\frac12 & & s'+\frac12 \\  & {s} &
\end{array} \right) &\xrightarrow{m=0}&\left( \begin{array}{c} {s+1}
\\ s+\frac12 \end{array}
\right)\oplus\sum_{k=1}^s \left( \begin{array}{ccc}
 & {k'+\frac{1}{2}} & \\ k & & k \\  & {k-\frac{1}{2}} &
\end{array} \right)\oplus\left( \begin{array}{c} {\frac{1}{2}'} \\ 0,0
\end{array} \right)
\end{eqnarray*}

Now let us consider 3D massive supermultiplets decomposition in the
massless limit and compare it with 4D case. First of all recall that
all 3D massless bosonic and fermionic higher spin fields do not
propagate any degrees of freedom. Only massless fields with spin $1,
1/2$ and $0$ propagate one physical degree of freedom. In turn 3D
massive higher spin fields do propagate physical degrees of freedom,
two and one d.o.f. for bosons and fermions respectively. In gauge the
invariant formulation this is clearly seen from (\ref{FD}).

As in 4D the minimal massless supermultiplets in 3D contain one
boson and one fermion. However unlike 4D massless higher spin fields
in 3D do not have physical degrees of freedom that is why one can
extend massless supermultiplet adding such fields. In some sense it
plays analogous role as the mixing between two massless
supermultiplets in 4D case (\ref{mix}). As we will see in 3D one can
construct extended massless supermultiplet which will correspond to
the massless decomposition of the massive supermultiplet.

Further in this section we first of all construct minimal massless
higher spin supermultiplets. They will play the role of initial blocks
for the construction of the extended massless supermultiplet. Then we
fined a gauge invariant massive deformation for them so that the
resulting system describe massive supermultiplets.

\subsection{Massless higher spin supermultiplets}

{\bf Supermultiplet $(k+\frac{3}{2},k+1)$, $k \ge 1$}. It contains
one fermionic field with spin $(k+\frac{3}{2})$ and one bosonic
field with spin $(k+1)$. In the frame-like formulation corresponding
field variables are the one 1-form $\Phi^{\alpha(2k+1)}$ for the
fermion and two 1-forms $\Omega^{\alpha(2k)}$, $f^{\alpha(2k)}$ for
boson. The Lagrangian describing this supermultiplet is just the sum
of their kinetic terms (see subsections \ref{massless_bosons} and
\ref{massless_fermions} for details):
\begin{eqnarray}
{\cal L}_0 &=& (-1)^{k+1} [ l \Omega_{\alpha(2k-1)\beta}
e^\beta{}_\gamma \Omega^{\alpha(2k-1)\gamma} + \Omega_{\alpha(2k)} d
f^{\alpha(2k)}  + \frac{i}{2} \Phi_{\alpha(2k+1)} d
\Phi^{\alpha(2k+1)}]
\end{eqnarray}
It is not hard to show that the Lagrangian is invariant under the
following global supertransformations:
\begin{eqnarray}
\delta f^{\alpha(2k)} = i(2k+1) \alpha_k \Phi^{\alpha(2k)\beta}
\zeta_\beta,\qquad \delta \Phi^{\alpha(2k+1)} = \alpha_k
\Omega^{\alpha(2k)} \zeta^\alpha
\end{eqnarray}
Let us calculate commutator of the supertransformation on the bosonic
field
\begin{eqnarray}
[\delta_1,\delta_2] f^{\alpha(2l)} &=&
i(2k+1)\alpha_k{}^2\Omega^{\alpha(2l-1)\beta}(\zeta_1{}_\beta\zeta_2{}^\alpha-\zeta_2{}_\beta\zeta_1{}^\alpha)
\end{eqnarray}
In the frame-like formulation the right side of the commutator
corresponds to the translations and it means that
$$
[Q_\alpha,Q_\beta]\sim P_{\alpha\beta}
$$
In what follows we will not fix the normalization of
supertransformations.
\\
{\bf Supermultiplet $(k+1,k+\frac12)$, $k \ge 1$}. It contains one
fermionic field with spin $(k+\frac{1}{2})$ and one bosonic field
with spin $(k+1)$. In the frame-like formulation the corresponding
field variables are one 1-form $\Phi^{\alpha(2k-1)}$ for fermion and
two 1-forms $\Omega^{\alpha(2k)}$, $f^{\alpha(2k)}$ for boson. The
Lagrangian describing this supermultiplets has the form:
\begin{eqnarray}
{\cal L}_0 &=& (-1)^{k+1} [ l \Omega_{\alpha(2k-1)\beta}
e^\beta{}_\gamma \Omega^{\alpha(2k-1)\gamma} + \Omega_{\alpha(2k)} d
f^{\alpha(2k)}  - \frac{i}{2} \Phi_{\alpha(2k-1)} d
\Phi^{\alpha(2k-1)}]
\end{eqnarray}
It is invariant under the following supertransformations
\begin{eqnarray}
\delta f^{\alpha(2k)} = i\beta_k \Phi^{\alpha(2l-1)}
\zeta^\alpha,\qquad \delta \Phi^{\alpha(2k-1)} = 2k\beta_k
\Omega^{\alpha(2k-1)\beta} \zeta_\beta
\end{eqnarray}

So we have described the full set of the massless higher spin
supermultiplets but in the massive case we will also need the massless
lower spin supermultiplets $(1,\frac{1}{2})$ and $(\frac{1}{2},0)$.
\\
{\bf Supermultiplet $(1,\frac{1}{2})$} contains the fermionic
zero-form $\psi^\alpha$ as well as the bosonic zero-form
$B^{\alpha\beta}$ and the one-form $A$. The sum of their kinetic terms
\begin{equation}\label{MLLS1}
{\cal L}_0 =  E B^{\alpha\beta} B_{\alpha\beta} - B_{\alpha\beta}
e^{\alpha\beta} d A + \frac{i}{2} \psi_\alpha E^\alpha{}_\beta d
\psi^\alpha
\end{equation}
One can show that on the auxiliary field $B^{\alpha\beta}$ equations
we have
\begin{eqnarray}\label{Low_eq1}
E^\alpha{}_\beta dB^{\beta\gamma}=E^\gamma{}_\beta dB^{\beta\alpha}
\end{eqnarray}
Lagrangian (\ref{MLLS1}) is invariant under the following
supertransformations\footnote{Strictly speaking this Lagrangian is
invariant up to the terms proportional to the auxiliary field
$B^{\alpha\beta}$ equation only (\ref{Low_eq1}). Thus there are two
possible approaches here. From one hand one can introduce non-trivial
corrections to the supertransformations for this auxiliary field.
Another possibility, that we will systematically follow here and
further on, is to use equations for the auxiliary fields in
calculating all variations.}:
\begin{eqnarray}
\delta A = i\beta_0 \phi_\alpha e^{\alpha\beta} \zeta_\beta,\qquad
\delta \phi^\alpha = 4\beta_0 B^{\alpha\beta} \zeta_\beta
\end{eqnarray}
{\bf Supermultiplet $(\frac{1}{2},0)$} contains the fermionic
zero-form $\phi_\alpha$ and two bosonic zero-forms $\pi^{\alpha\beta}$
and $\varphi$. The sum of their kinetic terms looks like:
\begin{equation}
{\cal L}_0 = \frac{i}{2} \phi_\alpha E^\alpha{}_\beta d \phi^\alpha
- E \pi^{\alpha\beta} \pi_{\alpha\beta} + E^{\alpha\beta}
\pi_{\alpha\beta} d \varphi
\end{equation}
Using the fact that on the auxiliary field $\pi^{\alpha\beta}$
equations we have
\begin{eqnarray}\label{Low_eq2}
E^\alpha{}_\gamma d \pi^{\beta\gamma}
=\frac12\varepsilon^{\alpha\beta}E_{\gamma\delta}d\pi^{\gamma\delta}
\end{eqnarray}
we can show that the Lagrangian is invariant under the following
supertransformations:
\begin{equation}
\delta \varphi = \frac{i\tilde{\delta}_0}{2} \phi_\alpha
\zeta^\alpha, \qquad \delta \phi^\alpha = \tilde{\delta}_0
\pi^{\alpha\beta} \zeta_\beta.
\end{equation}

\subsection{Massive higher spin supermultiplets}

{\bf Supermultiplet $(s+\frac{1}{2},s)$}
\\
This supermultiplet contains spin-$s+\frac{1}{2}$ fermion with mass
$m_1$ and spin-$s$ boson with mass $m$. In the massless limit it
decomposes into the system of the massless fields with spins
$$
(s+\frac12),s,(s-\frac12),(s-1),...,\frac32,1,\frac12,0
$$
Correspondingly we begin with the appropriate sum of the kinetic terms
for all fields
\begin{eqnarray}
{\cal L} &=& \sum_{k=1}^{s-1} (-1)^{k+1} [ k
\Omega_{\alpha(2k-1)\beta} e^\beta{}_\gamma
\Omega^{\alpha(2k-1)\gamma} + \Omega_{\alpha(2k)} d f^{\alpha(2k)}
 ] \nonumber \\
 && + E B_{\alpha\beta} B^{\alpha\beta} - B_{\alpha\beta}
e^{\alpha\beta} d A - E \pi_{\alpha\beta} \pi^{\alpha\beta} +
\pi_{\alpha\beta} E^{\alpha\beta} d \varphi \nonumber \\
 &&+\frac{i}{2}\sum_{k=0}^{s-1} (-1)^{k+1}
\Phi_{\alpha(2k+1)} d \Phi^{\alpha(2k+1)} + \frac{i}{2} \phi_\alpha
E^\alpha{}_\beta d \phi^\beta\label{lag0}
\end{eqnarray}
This Lagrangian possesses the following supersymmetry:
\begin{eqnarray}
\delta f^{\alpha(2k)} &=& i\beta_k\Phi^{\alpha(2k-1)}\zeta^\alpha
+i\alpha_k\Phi^{\alpha(2k)\beta}\zeta_\beta\nonumber
\\
\delta f^{\alpha(2)} &=&i\beta_1\Phi^{\alpha}\zeta^\alpha
+i\alpha_1\Phi^{\alpha(2)\beta}\zeta_\beta\nonumber
\\
\delta A &=&i\alpha_0\Phi^\alpha\zeta_\alpha
+i\beta_0e_{\alpha\beta}\phi^\alpha\zeta^\beta ,\qquad \delta\varphi
=-\frac{i\tilde\delta_0}{2}\phi^\gamma\zeta_\gamma\label{var0}
\\
\delta\Phi^{\alpha(2k+1)} &=&
\frac{\alpha_k}{(2k+1)}\Omega^{\alpha(2k)}\zeta^\alpha+
2(k+1)\beta_{k+1}\Omega^{\alpha(2k+1)\beta}\zeta_\beta\nonumber
\\
\delta\Phi^{\alpha} &=&
2\beta_1\Omega^{\alpha\beta}\zeta_\beta+\alpha_0e_{\beta(2)}B^{\beta(2)}\zeta^\alpha\nonumber
\\
\delta\phi^\alpha&=&{4\beta_0}B^{\alpha\beta}\zeta_\beta
+{\ti\delta_0}\pi^{\alpha\beta}\zeta_\beta\nonumber
\end{eqnarray}
These supertransformations are determined by the structure of the
supertransformations for the massless supermultiplets given above. So
in some sense the Lagrangian (\ref{lag0}) describes "large" massless
supermultiplet which contain the full set of the massless fields.

To construct a massive deformation we have to add the lower derivative
terms. For the bosonic terms we take the ones corresponding to the
gauge invariant description of massive spin-$s$ boson with mass $m$
(see subsection \ref{massive_boson})
\begin{eqnarray}
{\cal L}_b &=&  \sum_{k=1}^{s-2} (-1)^{k+1} a_k [ - \frac{(k+2)}{k}
\Omega_{\alpha(2)\beta(2k)} e^{\alpha(2)} f^{\beta(2k)} +
\Omega_{\alpha(2k)} e_{\beta(2)} f^{\alpha(2k)\beta(2)} ] \nonumber
\\
 && + 2a_0 \Omega_{\alpha(2)} e^{\alpha(2)} A - a_0
f_{\alpha\beta} E^\beta{}_\gamma B^{\alpha\gamma} + 2sM
\pi_{\alpha\beta} E^{\alpha\beta} A \nonumber \\
 && + \sum_{k=1}^{s-1} (-1)^{k+1} b_k f_{\alpha(2k-1)\beta}
e^\beta{}_\gamma f^{\alpha(2k-1)\gamma} + b_0 f_{\alpha(2)}
E^{\alpha(2)} \varphi + \frac{3a_0{}^2}{2}
 E \varphi^2 \label{lag2}
\end{eqnarray}
Note that after such deformation equations (\ref{Low_eq1}),
(\ref{Low_eq2}) are also deformed. Now they have the form
$$
E^\alpha{}_\gamma dB^{\beta\gamma}=E^\beta{}_\gamma
dB^{\alpha\gamma}-\frac{a_0}{4}\varepsilon^{\alpha\beta}e^{\gamma\delta}df_{\gamma\delta}
$$
$$
E^\alpha{}_\gamma
d\pi^{\beta\gamma}=\frac12\varepsilon^{\alpha\beta}E^{\gamma\delta}d\pi_{\gamma\delta}+
\frac{sM}{2}e^{\alpha\beta}dA
$$
For the fermionic terms we take the ones corresponding to the gauge
invariant description of the massive spin-$s+1/2$ fermion with mass
$m_1$ (see subsection \ref{massive_fermion})
\begin{eqnarray}
\frac{1}{i} {\cal L}_f &=& \sum_{k=1}^{s-1} (-1)^{k+1} c_k
\Phi_{\alpha(2k-1)\beta(2)} e^{\beta(2)} \Phi^{\alpha(2k-1)} + c_0
\Phi_\alpha E^\alpha{}_\beta
\phi^\beta \nonumber \\
 && + \sum_{k=0}^{s-1} (-1)^{k+1} \frac{d_k}{2}
\Phi_{\alpha(2k)\beta} e^\beta{}_\gamma \Phi^{\alpha(2k)\gamma} -
\frac{3d_0}{2} E \phi_\alpha \phi^\alpha \label{lagf}
\end{eqnarray}
Calculating the variations we use auxiliary field equations:
$$
e^\alpha{}_\gamma\Omega^{\alpha(2k-1)\gamma}=-df^{\alpha(2k)}-
\frac{a_{k-1}(k+1)}{k(k-1)(2k-1)}e^{\alpha(2)}f^{\alpha(2k-2)}-
a_ke_{\beta(2)}f^{\alpha(2k)\beta(2)}
$$
$$
EB^{\alpha(2)}=\frac12e^{\alpha(2)}dA-\frac{a_0}{4}E^{\alpha}{}_\gamma
f^{\alpha\gamma},\qquad
2E\pi^{\alpha(2)}=E^{\alpha(2)}d\varphi+2sME^{\alpha(2)}A
$$
In this case to find a massive deformation for the supertransformation
we have to add corrections for the fermionic fields only. From the
variations with one derivative we found that we must introduce the
following set of the corrections to the supertransformations:
\begin{eqnarray}\label{var1}
\delta\Phi^{\alpha(2k+1)} &=&\gamma_kf^{\alpha(2k)}\zeta^\alpha+
\delta_kf^{\alpha(2k+1)\beta}\zeta_\beta
\nonumber\\
\delta\Phi^{\alpha}
&=&\delta_0f^{\alpha\beta}\zeta_\beta+\gamma_0A\zeta^\alpha+
\tilde\gamma_0\varphi e^\alpha{}_\beta\zeta^\beta
\\
\delta\phi^\alpha&=&\rho_0\varphi\zeta^\alpha \nonumber
\end{eqnarray}
Taking into account that all the coefficients in the Lagrangian are
fixed, it is straightforward to find solution for the parameters of
the supertransformations which appears to be valid when $m_1=m$ only
\begin{eqnarray}\label{solution1}
\alpha_{k}{}^2 &=& k(s+k+1) \hat\alpha^2,\qquad \beta_{k}{}^2=
\frac{(k+1)(s-k)}{2k(2k+1)} \hat\alpha^2 \nonumber \\
\gamma_{k}{}^2&=&
\frac{s^2(s+k+1)}{4k(k+1)^2(2k+1)^2}m^2\hat\alpha^2
\\
\delta_{k}{}^2&=&\frac{s^2(s-k-1)}{2(k+1)(k+2)(2k+3)}
m^2\hat\alpha^2 \nonumber
\end{eqnarray}
$$
\ti\delta_0=4\beta_0=\sqrt{2s}\hat\alpha,
\qquad\alpha_0=\frac12\sqrt{(s+1)}\hat\alpha,\qquad
\hat\alpha^2=\frac{\alpha_{s-1}{}^2}{2s(s-1)}
$$
$$
\gamma_0=-2\ti\gamma_0=s\sqrt{(s+1)}m\hat\alpha,\quad
\rho_0=-\frac{(s+1)\sqrt{s}m}{\sqrt2}\hat\alpha
$$
\\
{\bf Supermultiplet $(s,s-\frac{1}{2})$}
\\
We need the same set of fields as in the previous case except the
field $\Phi^{\alpha(2s-1)}$. So we take the same massless Lagrangian
(\ref{lag0}) with this field omitted and the same set of initial
supertransformations (\ref{var0}) where now $\alpha_{s-1}=0$. As far
as the low derivative terms, the bosonic terms again have the
same form (\ref{lag2}), while the fermionic terms correspond to the
gauge invariant description of the massive spin-$s-1/2$ fermion (it
coincide with (\ref{lagf}) where $\Phi^{\alpha(2s-1)}$ is omitted).
Since the structure of the bosonic and the fermionic terms is the same
as in previous case the correction to the supertransformations also
look the same that is
\begin{eqnarray}\label{var2}
\delta\Phi^{\alpha(2k+1)} &=&\gamma_kf^{\alpha(2k)}\zeta^\alpha+
\delta_kf^{\alpha(2k+1)\beta}\zeta_\beta
\nonumber\\
\delta\Phi^{\alpha}
&=&\delta_0f^{\alpha\beta}\zeta_\beta+\gamma_0A\zeta^\alpha+
\tilde\gamma_0\varphi e^\alpha{}_\beta\zeta^\beta
\\
\delta\phi^\alpha&=&\rho_0\varphi\zeta^\alpha \nonumber
\end{eqnarray}
Calculating all variations we obtain relation between masses $m=m_1$
and the following expressions for the parameters determining
supertransformations:
\begin{eqnarray}
\alpha_{k}{}^2 &=& k(s-k-1) \hat\alpha^2,\qquad  \beta_{k}{}^2=
\frac{(k+1)(s+k)}{2k(2k+1)} \hat\alpha^2 \nonumber \\
\gamma_{k}{}^2&=&
\frac{s^2(s-k-1)}{4k(k+1)^2(2k+1)^2}m^2\hat\alpha^2
\\
\delta_{k}{}^2&=&\frac{s^2(s+k+1)}{2(k+1)(k+2)(2k+3)}
m^2\hat\alpha^2 \nonumber
\end{eqnarray}
$$
\ti\delta_0=4\beta_0=\sqrt{2s}\hat\alpha,
\qquad\alpha_0=\frac12\sqrt{(s-1)}\hat\alpha,\qquad
\hat\alpha^2=\frac{\alpha_{s-2}{}^2}{(s-2)}
$$
$$
\gamma_0=-2\ti\gamma_0=s\sqrt{(s-1)}m\hat\alpha,\quad
\rho_0=-\frac{(s-1)\sqrt{s}}{\sqrt2}m\hat\alpha
$$

Thus we explicitly construct the Lagrangian description for the two
massive supermultiplets $(s+1/2,s)$ and $(s,s-1/2)$. Both of these
massive supermultiplets have two bosonic and one fermionic (left)
degrees of freedom i.e. possess (1,0) supersymmetry. One can construct
massive supermultiplets possessing (1,1) supersymmetry with equal
number of bosonic and fermionic degrees of freedom if one introduces
one more massive fermion (right). In our previous work \cite{BSZ15} we
have constructed the particular cases of such massive supermultiplets
when supertransformations take the form of diagonal superposition of
two (1,1) supertransformations.

\section{Higher spin supermultiplets in 3D $AdS$ space}

In this section we study ${\cal N}=1$ supersymmetric construction
for the higher spin models in 3D $AdS$ space. As it have been shown by
\cite{AT86} ${\cal N}$-extended supersymmetry in 3D $AdS$ space exists
in several incarnations. They are so-called $(p,q)$ supersymmetries
where $p,q$ are integers and ${\cal N}=p+q$. The simplest $(1,0)$,
which we restrict ourselves with, is naturally associated with 3D
$AdS$ supergroup
$$
OSp(1,2) \otimes Sp(2)
$$
so that we have supersymmetry in the left sector only. In practice
this means that the massive higher spin supermultiplets as well as the
massless ones contain only one bosonic and one fermionic degrees of
freedom.

As in the previous section our construction is based on the gauge
invariant description of the higher spin fields but in contrast to
Minkowski space we work in terms of the gauge invariant curvatures
and the Lagrangians in the form (\ref{LagC_s}) and (\ref{LagC_s/2}).
Below we present a general scheme of our higher spin supermultiplet
construction and apply it to the massless and massive cases. At the
end we demonstrate how 3D $AdS$ superalgebra is realized in our
construction. In particular we show that the higher spin
supertransformations satisfy (1,0) superalgebra for which the
commutation relation of supercharges has the form
$$
\{Q_\alpha,Q_\beta\}\sim
P_{\alpha\beta}+\frac{\lambda}{2}M_{\alpha\beta}
$$
where $P_{\alpha(2)}$ and $M_{\alpha(2)}$ are the generators of 3D
$AdS$ algebra.

\subsection{Procedure of curvature deformation}

Here we present the general scheme of our higher spin supermultiplet
constructions in 3D based on the procedure of curvature deformation.
As it has been shown in section 2 and 3 the Lagrangian description of
the bosonic and fermionic higher spin fields can be reformulated in
terms of the gauge invariant curvatures. To supersymmetrize the system
of the bosonic and fermionic fields we covariantly deform the
corresponding curvatures by a background non-dynamical gravitino
1-form $\Psi^\alpha$ with global transformation in 3D $AdS$ being
$$
\delta\Psi^\alpha=D\zeta^\alpha+\frac{\lambda}{2}e^\alpha{}_\beta\zeta^\beta\quad
\Leftrightarrow\quad
D\zeta^\alpha=-\frac{\lambda}{2}e^\alpha{}_\beta\zeta^\beta
$$
where $\zeta^\alpha$ is a parameter of the supersymmetry. So such
construction can be interpreted as a supersymmetric theory in terms
of the background fields of the supergravity. Due to some difference
in the structure of the gauge invariant curvatures for the massless
and massive case we separately present the procedure of their
deformation.

{\bf Massless fields}

Massless higher spin bosonic and fermionic fields are described by
1-forms (see subsections \ref{massless_bosons} and
\ref{massless_fermions}) which we denote as $\Omega$ and $\Phi$
respectively. Let us also denote corresponding gauge invariant
curvatures as ${\cal{R}}$ and ${\cal{F}}$ which are 2-forms. They
 have the form (schematically)
$$
{\cal{R}}=D\Omega+(e\Omega),\qquad {\cal{F}}=D\Phi+(e\Phi)
$$
here $e\equiv e_\mu{}^{\alpha\beta}$ is a non-dynamical background
3D $AdS$ frame. The  Lagrangians then look as follows
\begin{eqnarray}
{\cal{L}}_{\Omega}=\Omega {\cal{R}},\qquad
{\cal{L}}_\Phi=\Phi{\cal{F}}
\end{eqnarray}
We note again that in the massless theory all curvatures are 2-form
and so in 3D $AdS$ there is no possibility to rewrite the Lagrangian
in terms of squares of them. It can be done in space-time dimensions
greater of equal to four.

At the first step in the supersymmetric construction we deform the
curvatures by the terms containing the background gravitino 1-form
$\Psi^\alpha$
$$
\Delta{\cal{R}}=(\Psi\Phi),\qquad \Delta{\cal{F}}=(\Psi\Omega)
$$
and require that the deformed curvatures transform covariantly, that
is
$$
\delta\hat{\cal{R}}=\delta({\cal{R}}+\Delta{\cal{R}})\sim{\cal{F}},
\qquad\delta\hat{\cal{F}}=\delta({\cal{F}}+\Delta{\cal{F}})\sim{\cal{R}}
$$
In turn, the covariant deformation of curvatures immediately defines
the form of the supertransformations
$$
\delta_\zeta\Omega=\frac{\delta(\Psi\Phi)}{\delta\Psi^\alpha}\zeta^\alpha,\qquad
\delta_\zeta\Phi=\frac{\delta(\Psi\Omega)}{\delta\Psi^\alpha}\zeta^\alpha
$$
Then supersymmetric Lagrangian is obtained by summing of
${\cal{L}}_{\Omega}$ and ${\cal{L}}_{\Phi}$ where the initial
curvatures are replaced by the deformed ones
\begin{eqnarray}
\hat{\cal{L}}=\Omega \hat{\cal{R}}+\Phi\hat{\cal{F}}
\end{eqnarray}

 {\bf Massive fields}

In the gauge invariant description of the massive bosonic and
fermionic higher spin fields the set of field variables contain
1-forms as well as 0-forms (see subsections \ref{massive_boson} and
\ref{massive_fermion}). We denote them respectively as $\Omega^A$,
$B^A$ for bosons and $\Phi^A$, $\phi^A$ for fermions. Let us also
denote the corresponding gauge invariant curvatures as ${\cal{R}}^A$,
${\cal{B}}^A$ and ${\cal{F}}^A$, ${\cal{C}}^A$ (${\cal{R}}^A$ and
${\cal{F}}^A$ are 2-forms, ${\cal{B}}^A$ and ${\cal{C}}^A$ are
1-forms). Schematically they have the following structure
\begin{align}
&{\cal{R}}^A=D\Omega^A+(e\Omega)^A+(eeB)^A &&
{\cal{F}}^A=D\Phi^A+(e\Phi)^A+(ee\phi)^A
\\
&{\cal{B}}^A=DB^A+\Omega+(eB)^A &&
{\cal{C}}^A=D\phi^A+\Phi+(e\phi)^A
\end{align}
Note that in 3D terms $(eeB)^A$ and $(ee\phi)^A$ in the expressions
for the 2-form curvatures disappear for higher spin components. It is
typical for 3D higher spin models and is related to the fact that the
massless higher spin fields do not propagate any degrees of freedom.
Unlike the massless case the Lagrangian for the massive higher spins
can be presented as the sum of the quadratic expressions in 2-form and
1-form curvatures (\ref{LagC_s}), (\ref{LagC_s/2})
\begin{eqnarray}
{\cal{L}}_\Omega=\sum {\cal{R}}^A {\cal{B}}^A,\qquad
{\cal{L}}_\Phi=\sum {\cal{F}}^A {\cal{C}}^A
\end{eqnarray}

Let us turn to a realization of the supersymmetric construction.
Deformation of the curvatures by the terms containing the background
gravitino 1-form $\Psi^\alpha$ schematically can be written as
\begin{align}
&\Delta{\cal{R}}^A=(\Psi\Phi)^A+(e\Psi\phi)^A &&
\Delta{\cal{F}}^A=(\Psi\Omega)^A+(e\Psi B)^A
\\
&\Delta{\cal{B}}^A=(\Psi\phi)^A && \Delta{\cal{C}}^A=(\Psi B)^A
\end{align}
Note here that the presence of $(e\Psi\phi)^A$ and $(e\Psi B)^A$ terms
in 2-form curvatures are relates with the presence of $(eeB)^A$ and
$(ee\phi)^A$ ones. Hence such terms in the deformations appear in the
2-form curvatures for the low spin components only. The requirement of
covariant deformation defines uniquely supertransformations
\begin{align}
&\delta_\zeta\Omega^A=(\frac{\delta(\Psi\Phi)^A}{\delta\Psi^\alpha}
+\frac{\delta(e\Psi\phi)^A}{\delta\Psi^\alpha})\zeta^\alpha &&
\delta_\zeta\Phi^A=(\frac{\delta(\Psi\Omega)^A}{\delta\Psi^\alpha}
+\frac{\delta(e\Psi B)^A}{\delta\Psi^\alpha})\zeta^\alpha
\\
&\delta_\zeta B^A=\frac{\delta(\Psi\phi)^A}{\Psi^\alpha}\zeta^\alpha
&& \delta\phi^A=\frac{\delta(\Psi B)^A}{\Psi^\alpha}\zeta^\alpha
\end{align}
Finally  the supersymmetric Lagrangian for the given supermultiplet is
the sum of the free Lagrangians where initial curvatures are replaced
by the deformed ones
\begin{eqnarray}
\hat{\cal{L}}=\sum[ \hat{\cal{R}}^A \hat{\cal{B}}^A+ \hat{\cal{F}}^A
\hat{\cal{C}}^A]
\end{eqnarray}
Possible arbitrariness is fixed by the condition that the Lagrangian
must be invariant under the supertransformations. Below we
demonstrate how such general scheme can actually be realized for the
massless and massive supermultiplets in 3D $AdS$.

\subsection{Massless supermultiplets}

{\bf Supermultiplet $(k+1,k+3/2)$}
\\
Integer spin-$(k+1)$ field is described by two 1-forms
$\Omega^{\alpha(2k)},f^{\alpha(2k)}$ and half-integer spin-$(k+3/2)$
is described by one 1-forms $\Phi^{\alpha(2k+1)}$. The initial
curvatures for these system are defined by (). Let us begin with the
deformation of the curvatures for the integer spin. There is a unique
possibility
\begin{eqnarray*}
\Delta{\cal{R}}^{\alpha(2k)} &=&
i\sigma_k\Phi^{\alpha(2k)\beta}\Psi_\beta
\\
\Delta{\cal{T}}^{\alpha(2k)} &=&
i\alpha_k\Phi^{\alpha(2k)\beta}\Psi_\beta
\end{eqnarray*}
where we have two arbitrary parameters $\sigma_k$ and $\alpha_k$. To
construct covariant deformation we write out the corresponding
supertransformations
\begin{eqnarray}\label{Super_b}
\delta\Omega^{\alpha(2k)} &=&
i\sigma_k\Phi^{\alpha(2k)\beta}\zeta_\beta
\nonumber\\
\delta f^{\alpha(2k)} &=& i\alpha_k\Phi^{\alpha(2k)\beta}\zeta_\beta
\end{eqnarray}
Explicit calculations of the covariant transformations for the
deformed curvatures give us, on one hand
\begin{eqnarray*}
\delta\hat{\cal{R}}^{\alpha(2k)}
&=&i\tilde\sigma_kD\Phi^{\alpha(2k)\beta}\zeta_\beta +
i\frac{\lambda^2}{4}\tilde\alpha_ke^\alpha{}_\gamma\Phi^{\alpha(2k-1)\gamma\beta}\zeta_\beta-
i\frac{\lambda}{2}\sigma_ke_\gamma{}_\beta\Phi^{\alpha(2k)\gamma}\zeta^\beta
\\
\delta\hat{\cal{T}}^{\alpha(2k)} &=&
i\tilde\alpha_kD\Phi^{\alpha(2k)\beta}\zeta_\beta+
i\tilde\sigma_ke^\alpha{}_\gamma\Phi^{\alpha(2k-1)\gamma\beta}\zeta_\beta-
i\frac{\lambda}{2}\alpha_ke_\gamma{}_\beta\Phi^{\alpha(2k)\gamma}\zeta^\beta
\end{eqnarray*}
and on the other hand
\begin{eqnarray*}
\delta\hat{\cal{R}}^{\alpha(2k)}
&=&i\tilde\sigma_k{\cal{F}}^{\alpha(2k)\beta}\zeta_\beta
\\
&=&i\tilde\sigma_k[D \Phi^{\alpha(2k)\beta} + \frac{\lambda}{2}(
e^\alpha{}_\gamma \Phi^{\alpha(2k-1)\beta\gamma}+e^\beta{}_\gamma
\Phi^{\alpha(2k)\gamma})]\zeta_\beta
\\
\delta\hat{\cal{T}}^{\alpha(2k)}
&=&i\tilde\alpha_k{\cal{F}}^{\alpha(2k)\beta}\zeta_\beta
\\
&=&i\tilde\alpha_k[D \Phi^{\alpha(2k)\beta} + \frac{\lambda}{2}(
e^\alpha{}_\gamma \Phi^{\alpha(2k-1)\beta\gamma}+e^\beta{}_\gamma
\Phi^{\alpha(2k)\gamma})]\zeta_\beta
\end{eqnarray*}
Comparing these expressions we obtain the relation on the parameters
$$
\sigma_k=\frac{\lambda}{2}\alpha_k
$$
Now let us turn to the curvature deformation for the half-integer spin
fields. A unique possibility is
\begin{eqnarray*}
\Delta{\cal{F}}^{\alpha(2k+1)}
&=&\beta_k\Omega^{\alpha(2k)}\Psi^\alpha+\gamma_kf^{\alpha(2k)}\Psi^\alpha
\end{eqnarray*}
where $\beta_k$ and $\gamma_k$ are another pair of arbitrary
parameters. The structure of the deformed curvatures defines the form
of the supertransformations
\begin{eqnarray}\label{Super_b}
\delta\Phi^{\alpha(2k+1)}
&=&\beta_k\Omega^{\alpha(2k)}\zeta^\alpha+\gamma_kf^{\alpha(2k)}\zeta^\alpha
\end{eqnarray}
Explicit verification of the covariant transformation for the deformed
curvatures gives on one hand
\begin{eqnarray*}
\delta\hat{\cal{F}}^{\alpha(2k+1)}
&=&\beta_kD\Omega^{\alpha(2k)}\zeta^\alpha+\gamma_kDf^{\alpha(2k)}\zeta^\alpha
+ \frac{\lambda}{2}
e^\alpha{}_\gamma[\beta_k\Omega^{\alpha(2k-1)\gamma}\zeta^\alpha
+\gamma_kf^{\alpha(2k-1)\gamma}\zeta^\alpha]
\end{eqnarray*}
and on the other hand
\begin{eqnarray*}
\delta\hat{\cal{F}}^{\alpha(2k+1)} &=&
\beta_k{\cal{R}}^{\alpha(2k)}\zeta^\alpha+\gamma_k{\cal{T}}^{\alpha(2k)}\zeta^\alpha\\
&=& \beta_k[D \Omega^{\alpha(2k)} + \frac{\lambda^2}{4}
e^\alpha{}_\beta f^{\alpha(2k-1)\beta}]\zeta^\alpha+\gamma_k[ D
f^{\alpha(2k)} + e^\alpha{}_\beta
\Omega^{\alpha(2k-1)\beta}]\zeta^\alpha
\end{eqnarray*}
Comparing these expressions we conclude that
$$
\gamma_k=\frac{\lambda}{2}\beta_k
$$
So we constructed the covariant supersymmetric deformation for the
curvatures but we still have two free parameters $\alpha_k$ and
$\beta_k$. To relate them we construct the supersymmetric Lagrangian.
We choose the following form for it
\begin{eqnarray}
{\cal L} &=&\frac{(-1)^{k+1}}{2}
[\Omega_{\alpha(2k)}\hat{\cal{T}}^{\alpha(2k)}+f_{\alpha(2k)}\hat{\cal{R}}^{\alpha(2k)}+
i \Phi_{\alpha(2k+1)}\hat{\cal{F}}^{\alpha(2k+1)}]
\end{eqnarray}
It is just sum of the free Lagrangians for spin-$(k+1)$ and
spin-$(k+3/2)$ fields where we have replaced curvatures. Using
supertransformations for fields and curvatures we obtain for the
Lagrangian variations
\begin{eqnarray*}
\delta{\cal L} &=&\frac{(-1)^{k+1}}{2}
[i(\sigma_k-(2k+1)\gamma_k)\Phi^{\alpha(2k)\beta}\zeta_\beta{\cal{T}}_{\alpha(2k)}
+i(\alpha_k-(2k+1)\beta_k)\Omega_{\alpha(2k)}{\cal{F}}^{\alpha(2k)\beta}\zeta_\beta\\
&&+i(\alpha_k-(2k+1)\beta_k)\Phi^{\alpha(2k)\beta}\zeta_\beta{\cal{R}}_{\alpha(2k)}
+i(\sigma_k-(2k+1)\gamma_k)f_{\alpha(2k)}{\cal{F}}^{\alpha(2k)\beta}\zeta_\beta]
\end{eqnarray*}
In order to achieve invariance we must put
$$
\alpha_k=(2k+1)\beta_k
$$
Thus we showed in the explicit form how using curvature deformation
procedure to construct supersymmetric Lagrangian formulation of the
multiplet $(k+1,k+3/2)$. Constructed model contain one free
parameter $\beta_k$ which is related with the normalization of the
superalgebra.
\\
{\bf Supermultiplet $(k+1,k+1/2)$}
\\
In this supermultiplet the integer spin is the same as in the previous
case and described by pair of 1-forms
$\Omega^{\alpha(2k)},f^{\alpha(2k)}$ and half-integer spin-$(k+1/2)$
is described by one 1-forms $\Phi^{\alpha(2k-1)}$. Supersymmetric
construction for the given supermultiplet is analogous and so let us
present the final results only. The supersymmetric curvatures have the
form
\begin{eqnarray*}
\hat{\cal{R}}^{\alpha(2k)} &=& D \Omega^{\alpha(2k)}
 + \frac{\lambda^2}{4} e^\alpha{}_\beta
f^{\alpha(2k-1)\beta}+i\frac{\lambda}{2}\beta_k\Phi^{\alpha(2k-1)}\Psi^\alpha
\nonumber \\
\hat{\cal{T}}^{\alpha(2k)} &=& D f^{\alpha(2k)} + e^\alpha{}_\beta
\Omega^{\alpha(2k-1)\beta}+i\beta_k\Phi^{\alpha(2k-1)}\Psi^\alpha
\\
\hat{\cal{F}}^{\alpha(2k-1)} &=& D \Phi^{\alpha(2k-1)} +
\frac{\lambda}{2} e^\alpha{}_\beta \Phi^{\alpha(2k-2)\beta}+
2k\beta_k(\Omega^{\alpha(2k-1)\beta}\Psi_\beta+
\frac{\lambda}{2}f^{\alpha(2k-1)\beta}\Psi_\beta)
\end{eqnarray*}
They are covariant under the following supertransformations
\begin{eqnarray*}
\delta\Omega^{\alpha(2k)} &=&
i\frac{\lambda}{2}\beta_k\Phi^{\alpha(2k-1)}\zeta^\alpha
\nonumber\\
\delta f^{\alpha(2k)} &=& i\beta_k\Phi^{\alpha(2k-1)}\zeta^\alpha
\\
\delta\Phi^{\alpha(2k-1)} &=&
2k\beta_k(\Omega^{\alpha(2k-1)\beta}\zeta_\beta+
\frac{\lambda}{2}f^{\alpha(2k-1)\beta}\zeta_\beta)
\end{eqnarray*}
The supersymmetric Lagrangian looks as follows
\begin{eqnarray}
{\cal L} &=&\frac{(-1)^{k+1}}{2}
[\Omega_{\alpha(2k)}\hat{\cal{T}}^{\alpha(2k)}+f_{\alpha(2k)}\hat{\cal{R}}^{\alpha(2k)}-
i \Phi_{\alpha(2k-1)}\hat{\cal{F}}^{\alpha(2k-1)}]
\end{eqnarray}
where
\begin{eqnarray*}
\delta\hat{\cal{R}}^{\alpha(2k)} &=&
i\frac{\lambda}{2}\beta_k{\cal{F}}^{\alpha(2k-1)}\zeta^\alpha
\nonumber\\
\delta \hat{\cal{T}}^{\alpha(2k)} &=&
i\beta_k{\cal{F}}^{\alpha(2k-1)}\zeta^\alpha
\\
\delta\hat{\cal{F}}^{\alpha(2k-1)} &=&
2k\beta_k({\cal{R}}^{\alpha(2k-1)\beta}\zeta_\beta+
\frac{\lambda}{2}{\cal{T}}^{\alpha(2k-1)\beta}\zeta_\beta)
\end{eqnarray*}

\subsection{Massive supermultiplets}

{\bf Supermultiplet $(s,s+\frac12)$}
\\
For the realization of the given massive supermultiplets let us first
consider their structure at the massless flat limit
$m,m_1,\lambda\rightarrow0$. In this case the Lagrangian is
described by the system of massless fields with spins
$(s+\frac12),s,...,\frac12,0$ in three dimensional flat space
\begin{eqnarray}
{\cal L} &=& \sum_{k=1}^{s-1} (-1)^{k+1} [ k
\Omega_{\alpha(2k-1)\beta} e^\beta{}_\gamma
\Omega^{\alpha(2k-1)\gamma} + \Omega_{\alpha(2k)} D f^{\alpha(2k)}
 ] \nonumber \\
 && + E B_{\alpha\beta} B^{\alpha\beta} - B_{\alpha\beta}
e^{\alpha\beta} D A - E \pi_{\alpha\beta} \pi^{\alpha\beta} +
\pi_{\alpha\beta} E^{\alpha\beta} D \varphi \nonumber \\
&&+\frac{i}{2}\sum_{k=0}^{s-1} (-1)^{k+1} \Phi_{\alpha(2k+1)} D
\Phi^{\alpha(2k+1)}  + \frac{1}{2} \phi_\alpha E^\alpha{}_\beta D
\phi^\beta
\end{eqnarray}
It is the same extended massless supermultiplet which we start the
construction of the massive supermultiplets in Minkowski space with. We
have shown that this Lagrangian is supersymmetric. If the equations
of motion (\ref{Low_eq1}), (\ref{Low_eq2}) are fulfilled, the
Lagrangian is invariant under the supertransformations (\ref{var0})
\begin{eqnarray*}
\delta f^{\alpha(2k)} &=& i\beta_k\Phi^{\alpha(2k-1)}\zeta^\alpha
+i\alpha_k\Phi^{\alpha(2k)\beta}\zeta_\beta
\\
\delta f^{\alpha(2)} &=&i\beta_1\Phi^{\alpha}\zeta^\alpha
+i\alpha_1\Phi^{\alpha(2)\beta}\zeta_\beta
\\
\delta A &=&i\alpha_0\Phi^\alpha\zeta_\alpha
+ic_0\beta_0e_{\alpha\beta}\phi^\alpha\zeta^\beta ,\qquad
\delta\varphi =-\frac{ic_0\tilde\delta_0}{2}\phi^\gamma\zeta_\gamma
\\
\delta\Phi^{\alpha(2k+1)} &=&
\frac{\alpha_k}{(2k+1)}\Omega^{\alpha(2k)}\zeta^\alpha+
2(k+1)\beta_{k+1}\Omega^{\alpha(2k+1)\beta}\zeta_\beta
\\
\delta\Phi^{\alpha} &=&
2\beta_1\Omega^{\alpha\beta}\zeta_\beta+2a_0\alpha_0e_{\beta(2)}B^{\beta(2)}\zeta^\alpha
\\
\delta\phi^\alpha&=&\frac{8a_0\beta_0}{c_0}B^{\alpha\beta}\zeta_\beta
+\frac{b_0\ti\delta_0}{c_0}\pi^{\alpha\beta}\zeta_\beta
\end{eqnarray*}
Here we take into account the normalization (\ref{Norm_s}),
(\ref{NormF}). Thus, requiring that massive theory has a correct
massless flat limit we partially fix an arbitrariness in the choice
of the supertransformations. Parameters
$\alpha_k,\beta_k,\beta_0,\alpha_0,\ti\delta_0$ at this step are
still arbitrary.

As in the massless case we will realize supersymmetry deforming the
curvatures by the background gravitino field $\Psi^\alpha$. We start
with the construction of the deformations for the bosonic fields
\begin{eqnarray*}
\Delta{\cal{R}}^{\alpha(2k)} &=&
i\rho_k\Phi^{\alpha(2k-1)}\Psi^\alpha+i\sigma_k\Phi^{\alpha(2k)\beta}\Psi_\beta
\\
\Delta{\cal{T}}^{\alpha(2k)} &=&
i\beta_k\Phi^{\alpha(2k-1)}\Psi^\alpha
+i\alpha_k\Phi^{\alpha(2k)\beta}\Psi_\beta
\\
\Delta{\cal{R}}^{\alpha(2)} &=& i\rho_1\Phi^{\alpha}\Psi^\alpha
+i\sigma_1\Phi^{\alpha(2)\beta}\Psi_\beta+i\hat\rho_0e^{\alpha(2)}\phi^\beta\Psi_\beta \\
\Delta{\cal{T}}^{\alpha(2)} &=&i\beta_1\Phi^{\alpha}\Psi^\alpha
+i\alpha_1\Phi^{\alpha(2)\beta}\Psi_\beta
\\
\Delta{\cal{A}} &=&i\alpha_0\Phi^\alpha\Psi_\alpha
+ic_0\beta_0e_{\alpha(2)}\phi^\alpha\Psi^\beta ,\qquad \Delta\Phi
=\frac{ic_0\tilde\delta_0}{2}\phi^\alpha\Psi_\alpha
\\
\Delta{\cal{B}}^{\alpha(2k)}
&=&-i\hat\rho_k\phi^{\alpha(2k-1)}\Psi^\alpha-i\hat\sigma_k\phi^{\alpha(2k)\beta}\Psi_\beta
\\
\Delta\Pi^{\alpha(2k)} &=&
-i\hat\beta_k\phi^{\alpha(2k-1)}\Psi^\alpha-i\hat\alpha_k\phi^{\alpha(2k)\beta}\Psi_\beta
\end{eqnarray*}
The corresponding ansatz for the supertransformations has the form
\begin{eqnarray}\label{Super_b}
\delta\Omega^{\alpha(2k)} &=&
i\rho_k\Phi^{\alpha(2k-1)}\zeta^\alpha+i\sigma_k\Phi^{\alpha(2k)\beta}\zeta_\beta
\nonumber\\
\delta f^{\alpha(2k)} &=& i\beta_k\Phi^{\alpha(2k-1)}\zeta^\alpha
+i\alpha_k\Phi^{\alpha(2k)\beta}\zeta_\beta
\nonumber\\
\delta\Omega^{\alpha(2)} &=& i\rho_1\Phi^{\alpha}\zeta^\alpha
+i\sigma_1\Phi^{\alpha(2)\beta}\zeta_\beta+i\hat\rho_0e^{\alpha(2)}\phi^\beta\zeta_\beta
\nonumber \\
\delta f^{\alpha(2)} &=&i\beta_1\Phi^{\alpha}\zeta^\alpha
+i\alpha_1\Phi^{\alpha(2)\beta}\zeta_\beta
\\
\delta A &=&i\alpha_0\Phi^\alpha\zeta_\alpha
+ic_0\beta_0e_{\alpha(2)}\phi^\alpha\zeta^\beta ,\qquad
\delta\varphi =-\frac{ic_0\tilde\delta_0}{2}\phi^\gamma\zeta_\gamma
\nonumber\\
\delta B^{\alpha(2k)}
&=&i\hat\rho_k\phi^{\alpha(2k-1)}\zeta^\alpha+i\hat\sigma_k\phi^{\alpha(2k)\beta}\zeta_\beta
\nonumber\\
\delta\pi^{\alpha(2k)} &=&
i\hat\beta_k\phi^{\alpha(2k-1)}\zeta^\alpha+i\hat\alpha_k\phi^{\alpha(2k)\beta}\zeta_\beta
\nonumber
\end{eqnarray}
All parameters are fixed by the requirement of covariant
transformations of the curvatures. First of all we consider
\begin{eqnarray}\label{CT_HS1}
\delta\hat{\cal{R}}^{\alpha(2k)} &=&
i\rho_k{\cal{F}}^{\alpha(2k-1)}\zeta^\alpha+i\sigma_k{\cal{F}}^{\alpha(2k)\beta}\zeta_\beta
\nonumber\\
\delta\hat{\cal{T}}^{\alpha(2k)} &=&
i\beta_k{\cal{F}}^{\alpha(2k-1)}\zeta^\alpha+i\alpha_k{\cal{F}}^{\alpha(2k)\beta}\zeta_\beta
\end{eqnarray}
It leads to relation $M_1=M+\frac{\lambda}{2}$ between mass
parameters $M_1$ and $M$ and defines the parameters
\begin{eqnarray}\label{solution1}
\alpha_{k}{}^2&=&k(s+k+1) [ M+(k+1) \lambda ]\hat\alpha^2
\nonumber\\
\beta_{k}{}^2&=& \frac{(k+1)(s-k)}{k(2k+1)} [ M-k\lambda
]\hat\beta^2
\nonumber\\
\sigma_{k}{}^2&=&\frac{(s+k+1)}{k(k+1)^2} [ M+(k+1) \lambda
]\hat\sigma^2
\nonumber\\
\rho_{k}{}^2&=&\frac{(s-k)}{k^3(k+1)(2k+1)} [ M-k\lambda]\hat\rho^2
\end{eqnarray}
where
$$
\hat\beta=\frac{\hat\alpha}{\sqrt2},\qquad
\hat\rho=\frac{sM}{2\sqrt2}\hat\alpha,\qquad
\hat\sigma=\frac{sM}{2}\hat\alpha,\qquad
\hat\alpha^2=\frac{\alpha_{s-1}{}^2}{2s(s-1)[ M+s \lambda ]}
$$
From the requirement that
\begin{eqnarray}\label{CT_HS2}
\delta\hat{\cal{B}}^{\alpha(2k)}&=&i\hat\rho_k{\cal{C}}^{\alpha(2k-1)}\zeta^\alpha
+i\hat\sigma_k{\cal{C}}^{\alpha(2k)\beta}\zeta_\beta
\nonumber\\
\delta\hat\Pi^{\alpha(2k)}&=&i\hat\beta_k{\cal{C}}^{\alpha(2k-1)}\zeta^\alpha
+i\hat\alpha_k{\cal{C}}^{\alpha(2k)\beta}\zeta_\beta
\end{eqnarray}
we obtain
$$
\hat\rho_k=\rho_k,\qquad
\hat\sigma_k=\sigma_k,\qquad\hat\beta_k=\beta_k,\qquad
\hat\alpha_k=\alpha_k
$$
The requirement of covariant transformations for the remaining
curvatures
\begin{eqnarray}\label{CT_HS3}
\delta\hat{\cal{R}}^{\alpha(2)} &=&
i\rho_1{\cal{F}}^{\alpha}\zeta^\alpha
+i\sigma_1{\cal{F}}^{\alpha(2)\beta}\zeta_\beta-i\hat\rho_0e^{\alpha(2)}{\cal{C}}^\beta\zeta_\beta
\nonumber\\
\delta\hat{\cal{T}}^{\alpha(2)} &=&
i\beta_1{\cal{F}}^{\alpha}\zeta^\alpha
+i\alpha_1{\cal{F}}^{\alpha(2)\beta}\zeta_\beta\nonumber\\
\delta\hat{\cal{A}} &=& i\alpha_0{\cal{F}}^\alpha\zeta_\alpha
-ic_0\beta_0e_{\alpha\beta}{\cal{C}}^\alpha\zeta^\beta,\qquad
\delta\hat\Phi
=-\frac{ic_0\tilde\delta_0}{2}{\cal{C}}^\gamma\zeta_\gamma
\end{eqnarray}
gives the solution
$$
\h\rho_0=-\frac18c_0{}^2\beta_1,\qquad\ti\delta_0=4\beta_0=\frac{c_0}{a_0}\beta_1,
\qquad\alpha_0=\frac{c_0{}^2}{4sMa_0}\beta_1
$$

Now let us consider the deformation of the curvatures for the
fermions. We choose an ansatz in the form
\begin{eqnarray*}
\Delta{\cal{F}}^{\alpha(2k+1)} &=&
\frac{\alpha_k}{(2k+1)}\Omega^{\alpha(2k)}\Psi^\alpha+
2(k+1)\beta_{k+1}\Omega^{\alpha(2k+1)\beta}\Psi_\beta\\
&& +\gamma_kf^{\alpha(2k)}\Psi^\alpha+
\delta_kf^{\alpha(2k+1)\beta}\Psi_\beta
\\
\Delta{\cal{F}}^{\alpha} &=&
2\beta_{1}\Omega^{\alpha\beta}\Psi_\beta+2a_0\alpha_0e_{\beta(2)}B^{\beta(2)}\Psi^\alpha
+ \delta_0f^{\alpha\beta}\Psi_\beta+\gamma_0A\Psi^\alpha+
\tilde\gamma_0\varphi e^\alpha{}_\beta\Psi^\beta
\\
\Delta{\cal{C}}^\alpha&=&-\frac{8a_0\beta_0}{c_0}B^{\alpha\beta}\Psi_\beta
-\frac{b_0\tilde\delta_0}{c_0}\pi^{\alpha\beta}\Psi_\beta
-\rho_0\varphi\Psi^\alpha
\\
\Delta{\cal{C}}^{\alpha(2k+1)}&=&-\tilde\beta_kB^{\alpha(2k+1)\beta}\Psi_\beta
-\tilde\alpha_k
B^{\alpha(2k)}\Psi^\alpha-\tilde\delta_k\pi^{\alpha(2k+1)\beta}\Psi_\beta
-\tilde\gamma_k\pi^{\alpha(2k)}\Psi^\alpha
\end{eqnarray*}
\\
and the ansatz for the supertransformations in the form
\begin{eqnarray}\label{Super_f}
\delta\Phi^{\alpha(2k+1)} &=&
\frac{\alpha_k}{(2k+1)}\Omega^{\alpha(2k)}\zeta^\alpha+
2(k+1)\beta_{k+1}\Omega^{\alpha(2k+1)\beta}\zeta_\beta\nonumber\\
&& +\gamma_kf^{\alpha(2k)}\zeta^\alpha+
\delta_kf^{\alpha(2k+1)\beta}\zeta_\beta
\nonumber\\
\delta\Phi^{\alpha} &=&
2\beta_{1}\Omega^{\alpha\beta}\zeta_\beta+2a_0\alpha_0e_{\beta(2)}B^{\beta(2)}\zeta^\alpha
+ \delta_0f^{\alpha\beta}\zeta_\beta+\gamma_0A\zeta^\alpha+
\tilde\gamma_0\varphi e^\alpha{}_\beta\zeta^\beta
\\
\delta\phi^\alpha&=&\frac{8a_0\beta_0}{c_0}B^{\alpha\beta}\zeta_\beta+
\frac{b_0\tilde\delta_0}{c_0}\pi^{\alpha\beta}\zeta_\beta
+\rho_0\varphi\zeta^\alpha
\nonumber\\
\delta\phi^{\alpha(2k+1)}&=&\tilde\beta_kB^{\alpha(2k+1)\beta}\zeta_\beta
+\tilde\alpha_k
B^{\alpha(2k)}\zeta^\alpha+\tilde\delta_k\pi^{\alpha(2k+1)\beta}\zeta_\beta
+\tilde\gamma_k\pi^{\alpha(2k)}\zeta^\alpha\nonumber
\end{eqnarray}

From the requirement that
\begin{eqnarray}\label{CT_HS4}
\delta\hat{\cal{F}}^{\alpha(2k+1)} &=&
\frac{\alpha_k}{(2k+1)}{\cal{R}}^{\alpha(2k)}\zeta^\alpha+
2(k+1)\beta_{k+1}{\cal{R}}^{\alpha(2k+1)\beta}\zeta_\beta\nonumber\\
&& +\gamma_k{\cal{T}}^{\alpha(2k)}\zeta^\alpha+
\delta_k{\cal{T}}^{\alpha(2k+1)\beta}\zeta_\beta
\end{eqnarray}
we have the same relation between masses $M_1=M+\frac{\lambda}{2}$.
Besides, it leads to
\begin{eqnarray*}
\gamma_{k}{}^2&=& \frac{(s+k+1)}{k(k+1)^2(2k+1)^2}[ M+(k+1) \lambda
]\hat\gamma^2
\\
\delta_{k}{}^2&=&\frac{(s-k-1)}{(k+1)(k+2)(2k+3)}[
M-(k+1)\lambda]\hat\delta^2
\end{eqnarray*}
where
$$
\hat\gamma=\frac{sM}{2}\hat\alpha,\qquad\hat\delta=\frac{sM}{\sqrt2}\hat\alpha
$$
In turn, the requirement that
\begin{eqnarray}\label{CT_HS5}
\delta\hat{\cal{C}}^{\alpha(2k+1)}&=&\tilde\beta_k{\cal{B}}^{\alpha(2k+1)\beta}\zeta_\beta
+\tilde\alpha_k
{\cal{B}}^{\alpha(2k)}\zeta^\alpha+\tilde\delta_k\Pi^{\alpha(2k+1)\beta}\zeta_\beta
+\tilde\gamma_k\Pi^{\alpha(2k)}\zeta^\alpha
\end{eqnarray}
gives us
$$
\tilde\gamma_k=\gamma_k,\quad\tilde\delta_k=\delta_k,\quad
\tilde\alpha_k=\frac{\alpha_k}{(2k+1)},\quad\tilde\beta_k=2(k+1)\beta_{k+1}
$$
At last, the requirement for the other curvatures
\begin{eqnarray}\label{CT_HS6}
\delta\hat{\cal{F}}^{\alpha}
&=&2\beta_{1}{\cal{R}}^{\alpha\beta}\zeta_\beta-2a_0\alpha_0e_{\beta(2)}{\cal{B}}^{\beta(2)}\zeta^\alpha
+
\delta_0{\cal{T}}^{\alpha\beta}\zeta_\beta+\gamma_0{\cal{A}}\zeta^\alpha+
\tilde\gamma_0\Phi e^\alpha{}_\beta\zeta^\beta
\nonumber\\
\delta\hat{\cal{C}}^\alpha&=&\frac{8a_0\beta_0}{c_0}{\cal{B}}^{\alpha\beta}\zeta_\beta
+\frac{b_0\ti\delta_0}{c_0}\Pi^{\alpha\beta}\zeta_\beta
+\rho_0\Phi\zeta^\alpha
\end{eqnarray}
yields solution
$$
\gamma_0=-2\ti\gamma_0=\frac{c_0{}^2}{2a_0}\beta_1,\quad
\rho_0=-\frac{c_0{}^2}{4sMa_0}\beta_1
$$
Now, all the arbitrary parameters are fixed.

The supersymmetric Lagrangian is the sum of the free Lagrangians where
the initial curvatures are replacement by the deformed ones
\begin{eqnarray}\label{LagSC_s}
\h{\cal{L}}&=&-\frac{1}{2}\sum_{k=1}^{s-1}(-1)^{k+1}[\h{\cal{R}}_{\alpha(2k)}\h\Pi^{\alpha(2k)}
+\h{\cal{T}}_{\alpha(2k)}\h{\cal{B}}^{\alpha(2k)}]
+\frac{a_0}{2sM}e_{\alpha(2)}\h{\cal{B}}^{\alpha(2)}\h\Phi\nonumber\\
&&-\frac{i}{2}\sum_{k=0}^{s-1}(-1)^{k+1}\h{\cal{F}}_{\alpha(2k+1)}\h{\cal{C}}^{\alpha(2k+1)}
\end{eqnarray}
The Lagrangian is invariant under the supertransformations for
curvatures up to equations of motion for the fields
$B^{\alpha(2)},\pi^{\alpha(2)}$
\begin{eqnarray}\label{Eq_low}
\Phi=0,\quad{\cal{A}}=0\quad\Rightarrow\quad
e_{\gamma(2)}\Pi^{\gamma(2)}=D\Phi-2sM{\cal{A}}=0
\end{eqnarray}

The Lagrangian (\ref{LagSC_s}) is a final solution for the massive
supermultiplet $(s,s+\frac12)$.
\\
{\bf Supermultiplet $(s,s-\frac12)$}
\\
In this section we consider another massive higher spin
supermultiplet when the highest spin is boson. The massive spin-$s$
field was described in section 3.1 in terms of massless fields. The
massive spin-$(s-1/2)$ field can be obtained for the results in
section 3.2 if one makes the replacement $s\rightarrow(s-1)$. So the
set of massless fields for the massive field with spin $s-1/2$ is
$\Phi^{\alpha(2k+1)}$, $0 \le k \le s-2$ and $\phi^{\alpha}$. The
gauge invariant curvatures and the Lagrangian have the forms
(\ref{GT_s/2}) and (\ref{LagC_s/2}) where the parameters are
\begin{eqnarray}\label{InD}
c_k{}^2 &=& \dfrac{(s+k)(s-k-1)}{2(k+1)(2k+1)} [ M_1{}^2 - (2k+1)^2
\frac{\lambda^2}{4} ]
\nonumber\\
c_0{}^2 &=& 2s(s-1) [ M_1{}^2 - \frac{\lambda^2}{4} ]
\\
d_k &=& \dfrac{(2s-1)}{(2k+3)} M_1, \qquad M_1{}^2 = m_1{}^2 +
(s-\frac{3}{2})^2 \lambda^2\nonumber
\end{eqnarray}

Following our procedure we should construct the supersymmetric
deformations for the curvatures. Actually the structure of the
deformed curvatures and supertransformations have the same form as in
previous subsection for the supermultiplets $(s,s+1/2)$. There is a
difference in parameters (\ref{InD}) only. Therefore we present here
only the supertransformations for the curvatures. The requirement of
covariant curvature transformations for the bosonic fields
\begin{eqnarray*}
\delta\hat{\cal{R}}^{\alpha(2k)} &=&
i\rho_k{\cal{F}}^{\alpha(2k-1)}\zeta^\alpha+i\sigma_k{\cal{F}}^{\alpha(2k)\beta}\zeta_\beta
\\
\delta\hat{\cal{T}}^{\alpha(2k)} &=&
i\beta_k{\cal{F}}^{\alpha(2k-1)}\zeta^\alpha+i\alpha_k{\cal{F}}^{\alpha(2k)\beta}\zeta_\beta
\\
\delta\hat{\cal{R}}^{\alpha(2)} &=&
i\rho_1{\cal{F}}^{\alpha}\zeta^\alpha
+i\sigma_1{\cal{F}}^{\alpha(2)\beta}\zeta_\beta-i\hat\rho_0e^{\alpha(2)}{\cal{C}}^\beta\zeta_\beta
\\
\delta\hat{\cal{T}}^{\alpha(2)} &=&
i\beta_1{\cal{F}}^{\alpha}\zeta^\alpha
+i\alpha_1{\cal{F}}^{\alpha(2)\beta}\zeta_\beta\\
\delta\hat{\cal{A}} &=& i\alpha_0{\cal{F}}^\alpha\zeta_\alpha
-ic_0\beta_0e_{\alpha\beta}{\cal{C}}^\alpha\zeta^\beta,\qquad
\delta\hat\Phi
=-\frac{ic_0\tilde\delta_0}{2}{\cal{C}}^\gamma\zeta_\gamma
\\
\delta\hat{\cal{B}}^{\alpha(2k)}&=&i\hat\rho_k{\cal{C}}^{\alpha(2k-1)}\zeta^\alpha
+i\hat\sigma_k{\cal{C}}^{\alpha(2k)\beta}\zeta_\beta
\\
\delta\hat\Pi^{\alpha(2k)}&=&i\hat\beta_k{\cal{C}}^{\alpha(2k-1)}\zeta^\alpha
+i\hat\alpha_k{\cal{C}}^{\alpha(2k)\beta}\zeta_\beta
\end{eqnarray*}
gives us the relation $M_1=M-\frac{\lambda}{2}$ between masses $M_1$
and $M$.  Besides, it leads to
\begin{eqnarray}\label{SupMult2}
\sigma_{k}{}^2&=&\frac{(s-k-1)}{k(k+1)^2} [ M-(k+1) \lambda
]\hat\sigma^2\nonumber \\
\rho_{k}{}^2&=&\frac{(s+k)}{k^3(k+1)(2k+1)} [ M+k\lambda]\hat\rho^2
\nonumber\\
\alpha_{k}{}^2&=&k(s-k-1) [ M-(k+1) \lambda ]\hat\alpha^2
\nonumber\\
\beta_{k}{}^2&=& \frac{(k+1)(s+k)}{k(2k+1)} [ M+k\lambda
]\hat\beta^2
\end{eqnarray}
$$
\hat\rho_k=\rho_k,\qquad
\hat\sigma_k=\sigma_k,\qquad\hat\beta_k=\beta_k,\qquad
\hat\alpha_k=\alpha_k
$$
$$
\h\rho_0=-\frac18c_0{}^2\beta_1,\qquad\ti\delta_0=4\beta_0=\frac{c_0}{a_0}\beta_1,
\qquad\alpha_0=\frac{c_0{}^2}{4sMa_0}\beta_1
$$
where
$$
\hat\beta=\frac{\hat\alpha}{\sqrt2},\qquad
\hat\rho=\frac{sM}{2\sqrt2}\hat\alpha,\qquad
\hat\sigma=\frac{sM}{2}\hat\alpha,\qquad
\hat\alpha^2=\frac{\alpha_{s-2}{}^2}{(s-2)[ M-(s-1) \lambda ]}
$$
From the requirement of covariant supertransformations for the
fermionic curvatures
\begin{eqnarray*}
\delta\hat{\cal{F}}^{\alpha(2k+1)} &=&
\frac{\alpha_k}{(2k+1)}{\cal{R}}^{\alpha(2k)}\zeta^\alpha+
2(k+1)\beta_{k+1}{\cal{R}}^{\alpha(2k+1)\beta}\zeta_\beta\\
&& +\gamma_k{\cal{T}}^{\alpha(2k)}\zeta^\alpha+
\delta_k{\cal{T}}^{\alpha(2k+1)\beta}\zeta_\beta
\\
\delta\hat{\cal{F}}^{\alpha}
&=&2\beta_{1}{\cal{R}}^{\alpha\beta}\zeta_\beta-2a_0\alpha_0e_{\beta(2)}{\cal{B}}^{\beta(2)}\zeta^\alpha
+
\delta_0{\cal{T}}^{\alpha\beta}\zeta_\beta+\gamma_0{\cal{A}}\zeta^\alpha+
\tilde\gamma_0\Phi e^\alpha{}_\beta\zeta^\beta
\\
\delta\hat{\cal{C}}^\alpha&=&\frac{8a_0\beta_0}{c_0}{\cal{B}}^{\alpha\beta}\zeta_\beta
+\frac{b_0\ti\delta_0}{c_0}\Pi^{\alpha\beta}\zeta_\beta
+\rho_0\Phi\zeta^\alpha\\
\delta\hat{\cal{C}}^{\alpha(2k+1)}&=&\tilde\beta_k{\cal{B}}^{\alpha(2k+1)\beta}\zeta_\beta
+\tilde\alpha_k
{\cal{B}}^{\alpha(2k)}\zeta^\alpha+\tilde\delta_k\Pi^{\alpha(2k+1)\beta}\zeta_\beta
+\tilde\gamma_k\Pi^{\alpha(2k)}\zeta^\alpha\\
\end{eqnarray*}
one gets
\begin{eqnarray*}
\gamma_{k}{}^2&=& \frac{(s-k-1)}{k(k+1)^2(2k+1)^2}[ M-(k+1) \lambda
]\hat\gamma^2
\\
\delta_{k}{}^2&=&\frac{(s+k+1)}{(k+1)(k+2)(2k+3)}[
M+(k+1)\lambda]\hat\delta^2
\end{eqnarray*}
$$
\tilde\gamma_k=\gamma_k,\quad\tilde\delta_k=\delta_k,\quad
\tilde\alpha_k=\frac{\alpha_k}{(2k+1)},\quad\tilde\beta_k=2(k+1)\beta_{k+1}
$$
$$
\gamma_0=-2\ti\gamma_0=\frac{c_0{}^2}{2a_0}\beta_1,\quad
\rho_0=-\frac{c_0{}^2}{4sMa_0}\beta_1
$$
where
$$
\hat\gamma=\frac{sM}{2}\hat\alpha,\qquad\hat\delta=\frac{sM}{\sqrt2}\hat\alpha
$$
Supersymmetric Lagrangian have the form (\ref{LagSC_s}) and it is
invariant under the supertransformations up to equations of motion
for the auxiliary fields $B^{\alpha(2)},\pi^{\alpha(2)}$
(\ref{Eq_low}).

\subsection{Realization of (super)algebra}

In this section we analyze the commutators of the
(super)transformations and show how the (super)algebra is realized in
our construction. All the considerations are valid both for the
$(s,s+1/2)$ supermultiplets and the for $(s,s-1/2)$ one.

\subsubsection*{ Description of the $AdS$ transformations}

Before we turn to the supersymmetric theory let us discuss the
conventional $AdS_3$ algebra. In the frame formalism, $AdS$ space is
described by the background Lorentz connection field
$\omega^{\alpha(2)}$ and the background frame field $e^{\alpha(2)}$.
First of them enters implicitly through the covariant derivative
$D$, while the second one enters explicitly. Let $\eta^{\alpha(2)}$
and $\xi^{\alpha(2)}$ be the parameters of the Lorentz transformations
and the pseudo-translations respectively. The theory of the massive
spin-$s$ field has the following laws under these transformations
\begin{eqnarray}\label{Lorentz_b1}
\delta_\eta\Omega^{\alpha(2k)}=\eta^\alpha{}_\beta\Omega^{\alpha(2k-1)\beta}\qquad
\delta_\eta f^{\alpha(2k)}=\eta^\alpha{}_\beta f^{\alpha(2k-1)\beta}
\end{eqnarray}
\begin{eqnarray}\label{Translaton_b1}
\delta_\xi\Omega^{\alpha(2k)} &=& \frac{(k+2)a_k}{k} \xi_{\beta(2)}
\Omega^{\alpha(2k)\beta(2)} + \frac{a_{k-1}}{k(2k-1)}
\xi^{\alpha(2)} \Omega^{\alpha(2k-2)}
 + \frac{b_k}{k} \xi^\alpha{}_\beta f^{\alpha(2k-1)\beta} \nonumber \\
\delta_\xi f^{\alpha(2k)} &=& \xi^\alpha{}_\beta
\Omega^{\alpha(2k-1)\beta} + a_k \xi_{\beta(2)}
f^{\alpha(2k)\beta(2)} + \frac{(k+1)a_{k-1}}{k(k-1)(2k-1)}
\xi^{\alpha(2)} f^{\alpha(2k-2)}
\end{eqnarray}
here $a_k$ and $b_k$ are defined by (\ref{boson_data}). For the
massive spin-$(s\pm1/2)$ field the transformation laws look like
\begin{eqnarray}\label{ADStr_f}
\delta_\eta\Phi^{\alpha(2k+1)} &=&
\eta^\alpha{}_\beta\Phi^{\alpha(2k)\beta}
\nonumber \\
\delta_\xi\Phi^{\alpha(2k+1)} &=& \frac{d_k}{(2k+1)}
\xi^\alpha{}_\beta \Phi^{\alpha(2k)\beta}  + \frac{c_k}{k(2k+1)}
\xi^{\alpha(2)} \Phi^{\alpha(2k-1)}\nonumber\\
&&
 + c_{k+1} \xi_{\beta(2)} \Phi^{\alpha(2k+1)\beta(2)}
\end{eqnarray}
here $c_k$ and $d_k$ are defined by (\ref{fermion_data}) for the
spin-$(s+1/2)$ and (\ref{InD}) for the spin-$(s-1/2)$. To consider a
structure of the $AdS_3$ algebra $Sp(2) \otimes Sp(2)$ in the left
sector only we introduce the new variables for the bosonic fields
\begin{eqnarray}\label{NV}
\hat\Omega^{\alpha(2k)}=\Omega^{\alpha(2k)}+\frac{sM}{2k(k+1)}f^{\alpha(2k)},\qquad
\hat
f^{\alpha(2k)}=\Omega^{\alpha(2k)}-\frac{sM}{2k(k+1)}f^{\alpha(2k)}
\end{eqnarray}
so that the variables $\hat\Omega^{\alpha(2k)}$ correspond to the left
sector. In terms of these variables the transformations
(\ref{Lorentz_b1}), (\ref{Translaton_b1}) have the form
\begin{eqnarray}\label{ADStr_b}
\delta_\eta\hat\Omega^{\alpha(2k)}&=&\eta^\alpha{}_\beta\hat\Omega^{\alpha(2k-1)\beta}
\nonumber\\
\delta_\xi\hat\Omega^{\alpha(2k)} &=& \frac{(k+2)a_k}{k}
\xi_{\beta(2)} \hat\Omega^{\alpha(2k)\beta(2)} +
\frac{a_{k-1}}{k(2k-1)} \xi^{\alpha(2)}
\hat\Omega^{\alpha(2k-2)}\nonumber\\
&&
 + \frac{sM}{2k(k+1)} \xi^\alpha{}_\beta
\hat\Omega^{\alpha(2k-1)\beta}
\end{eqnarray}
Now let us consider the commutators of these transformations. The
direct calculations leads to the following results
\begin{eqnarray*}
{[\delta_{\eta_1},\delta_{\eta_2}]}\hat\Omega^{\alpha(2k)}&=&(\eta_2{}^\alpha{}_\beta\eta_1{}^\beta{}_\gamma-
\eta_1{}^\alpha{}_\beta\eta_2{}^\beta{}_\gamma)\hat\Omega^{\alpha(2k-1)\gamma},
\\
{[\delta_{\xi_1},\delta_{\xi_2}]}\hat\Omega^{\alpha(2k)}&=&
\frac{\lambda^2}{4} (\xi_2{}^\alpha{}_\gamma\xi_1{}^\gamma{}_\beta-
\xi_1{}^\alpha{}_\gamma\xi_2{}^\gamma{}_\beta)
\hat\Omega^{\alpha(2k-1)\beta},
\\
{[\delta_\eta,\delta_\xi]}\hat\Omega^{\alpha(2k)}&=&2\frac{(k+2)a_k}{k}
\xi_{\beta(2)}\eta^\beta{}_\gamma\hat\Omega^{\alpha(2k)\beta\gamma}
 + \frac{a_{k-1}}{k(2k-1)}\xi^{\alpha}{}_{\gamma}\eta^\alpha{}^\gamma
\hat\Omega^{\alpha(2k-2)}\\
&&  + \frac{sM}{2k(k+1)} (\xi^\alpha{}_\beta
\eta^\beta{}_\gamma-\eta^\alpha{}_\gamma\xi^\gamma{}_\beta)\hat\Omega^{\alpha(2k-1)}
\end{eqnarray*}
Comparison with (\ref{ADStr_b}) shows that we do have the
$AdS$-algebra
$$
[M_{\alpha(2)},M_{\beta(2)}]\sim\varepsilon_{\alpha\beta}M_{\alpha\beta}
,\qquad
[P_{\alpha(2)},P_{\beta(2)}]\sim\lambda^2\varepsilon_{\alpha\beta}M_{\alpha\beta}
$$
$$
[M_{\alpha(2)},P_{\beta(2)}]\sim\varepsilon_{\alpha\beta}P_{\alpha\beta}
$$
The analogous results have place for the commutators in the fermionic
sector as well.

\subsubsection*{ $AdS$ supertransformations}

Let us consider the supersymmetric theory.  The supertransformations
for the massive higher spin supermultiplets have the form
(\ref{Super_b}), (\ref{Super_f})
\begin{eqnarray*}
\delta\Omega^{\alpha(2k)} &=&
\frac{isM}{2k(k+1)}\beta_k\Phi^{\alpha(2k-1)}\zeta^\alpha+\frac{isM}{2k(k+1)}\alpha_k\Phi^{\alpha(2k)\beta}\zeta_\beta
\\
\delta f^{\alpha(2k)} &=& i\beta_k\Phi^{\alpha(2k-1)}\zeta^\alpha
+i\alpha_k\Phi^{\alpha(2k)\beta}\zeta_\beta
\\
\delta\Phi^{\alpha(2k+1)} &=&
\frac{\alpha_k}{(2k+1)}\Omega^{\alpha(2k)}\zeta^\alpha+
2(k+1)\beta_{k+1}\Omega^{\alpha(2k+1)\beta}\zeta_\beta\\
&& +\frac{sM}{2k(k+1)(2k+1)}\alpha_kf^{\alpha(2k)}\zeta^\alpha+
\frac{sM}{(k+2)}\beta_{k+1}f^{\alpha(2k+1)\beta}\zeta_\beta
\end{eqnarray*}
where the parameters $\alpha_k$ and $\beta_k$ are defined by
(\ref{solution1}) for the $(s,s+1/2)$ supermultiplets and
(\ref{SupMult2}) for the $(s,s-1/2)$ one. In terms of the new
variables (\ref{NV}) the supertransformations look like
\begin{eqnarray*}
\delta\hat\Omega^{\alpha(2k)} &=&
\frac{isM}{k(k+1)}\beta_k\Phi^{\alpha(2k-1)}\zeta^\alpha+\frac{isM}{k(k+1)}\alpha_k\Phi^{\alpha(2k)\beta}\zeta_\beta
\\
\delta \hat f^{\alpha(2k)} &=&0
\\
\delta\Phi^{\alpha(2k+1)} &=&
\frac{\alpha_k}{(2k+1)}\hat\Omega^{\alpha(2k)}\zeta^\alpha+
2(k+1)\beta_{k+1}\hat\Omega^{\alpha(2k+1)\beta}\zeta_\beta
\end{eqnarray*}
One can see that the $\hat f^{\alpha(2k)}$ fields are inert under
the  supertransformations. It just means that we have (1,0)
supersymmetry. Let us calculate the commutator of two
supertransformations. We obtain
\begin{eqnarray*}
[\delta_1,\delta_2]\hat\Omega^{\alpha(2k)}&=&isM\hat\alpha^2[
\frac{a_{k-1}}{k(2k-1)}\hat\Omega^{\alpha(2k-1)}\zeta_1{}^\alpha\zeta_2{}^\alpha+
\frac{2(k+2)a_k}{k}\hat\Omega^{\alpha(2k)\gamma\beta}\zeta_1{}_\beta\zeta_2{}_\gamma\\
&&
+\frac{sM}{k(k+1)}\hat\Omega^{\alpha(2k-1)\gamma}\zeta_1{}^\alpha\zeta_2{}_\gamma
+\lambda\hat\Omega^{\alpha(2k-1)\gamma}\zeta_1{}^\alpha\zeta_2{}_\gamma]
-(1\leftrightarrow2)
\end{eqnarray*}
\begin{eqnarray*}
[\delta_1,\delta_2]\hat\Phi^{\alpha(2k+1)} &=&
isM\hat\alpha^2[\frac{c_k}{k(2k+1)}\Phi^{\alpha(2k-1)}\zeta_1{}^\alpha\zeta_2{}^\alpha
+2c_{k+1}\Phi^{\alpha(2k+1)\gamma\beta}\zeta_1{}_\beta\zeta_2{}_\gamma\\
&&+
\frac{2d_k}{(2k+1)}\Phi^{\alpha(2k)\gamma}\zeta_1^\alpha\zeta_2{}_\gamma
+\lambda\Phi^{\alpha(2k)\gamma}\zeta_1^\alpha\zeta_2{}_\gamma ]
-(1\leftrightarrow2)
\end{eqnarray*}
Here we use the explicit expressions for $\alpha_k$ and $\beta_k$
and the conditions
$$
\frac{2\beta_k{}^2}{(k+1)}+\frac{\alpha_k{}^2}{k(k+1)(2k+1)}=\hat\alpha^2[\frac{sM}{k(k+1)}+\lambda]
$$
$$
\frac{2\beta_{k+1}{}^2}{(k+2)}+\frac{\alpha_k{}^2}{k(k+1)(2k+1)}=
\hat\alpha^2[\frac{2d_k}{(2k+1)}+\lambda]
$$
$$
\alpha_{k-1}\beta_k=(k+1)\hat\alpha^2a_{k-1},\quad
\alpha_k\beta_{k+1}=(k+2)\hat\alpha^2a_k,\quad
\alpha_k\beta_k=(k+1)c_k\hat\alpha^2
$$
Comparing the commutators of the supertransformations with
(\ref{ADStr_f}) we obtain the (1,0) $AdS_3$ superalgebra with the
commutation relation
\begin{eqnarray}\label{algebra}
\{Q_\alpha,Q_\beta\}\sim
P_{\alpha\beta}+\frac{\lambda}{2}M_{\alpha\beta}
\end{eqnarray}

As we see, the algebra of the supertransformations is closed. It is
worth emphasizing that we did not apply the equations of motion to
obtain the relation (\ref{algebra}) both in the bosonic and in the
fermionic sectors. This situation is analogous to the one for the
massless higher-spin fields in the three-dimensional frame-like
formalism. Recall that in the massive supermultiplet cases the
invariance of the Lagrangians is achieved up to the terms proportional
to the spin-1 and spin-0 auxiliary fields equations only. Note that in
dimensions $d \ge 4$ one would have to use equations for the higher
spins auxiliary fields as well (though in odd dimensions there exist
examples of the theories where Lagrangians are invariant without any
use of e.o.m \cite{Zan05}). The difference here comes from the well
known fact that all massless higher spin fields in three dimensions do
not have any local degrees of freedom.

%%%%%%%%%%%%%%%%%%%%%%%%%%%%%%%%%%%%%%%%%%%%%%%%%%%%%%%%%%%%%%%%%%%
\section{Summary}

In this review we have presented and discussed the component
supersymmetric formulations of the higher spin fields in three
dimensions. We applied these formulations for the Lagrangian
construction of the on-shell ${\cal N}=1$ massless and massive
supermultiplets in 3D Minkowski and AdS spaces. Although the
off-shell formulation of 3D massless higher spin supermultiplets is
known long enough we discussed the on-shell massless formulation as
well since it is interesting by itself. Besides such a formulation is
a base for the deformation to the massive higher spin supermultiplets.
To generate the massive terms in the Lagrangians we have used the
approach based on the gauge invariant formulation of the massive
higher spin fields where the massive fields are described as the
system of the massless ones coupled to each other in a special way.

In 3D Minkowski space we generalized the gauge invariant formulation
of massive higher spin fields to the case of the massive
supermultiplets. In particular we showed that the massive
supermultiplets can be constructed from the extended massless one.
This extended massless supermultiplet is defined and its smooth mass
deformation is explicitly constructed for the two case of the
massive supermultiplets $(s,s+1/2)$ and the $(s,s-1/2)$. Both of these
massive supermultiplets have two bosonic and one fermionic (left)
degrees of freedom, i.e. possess (1,0) supersymmetry.

In 3D AdS space the construction of the ${\cal N}=1$ massive
supermultiplets with (1,0) supersymmetry is realized by another more
elegant way. We have used a technique of the gauge invariant
curvatures. Namely, we found their supersymmetric deformation by the
background gravitino field. It means that resulting theory describes
the massive supermultiplets on the background of $AdS$ supergravity.
Constructed higher spin supermultiplets have a correct flat limit.

\section*{Acknowledgments}
I.L.B and T.V.S are grateful to the RFBR grant, project No.
15-02-03594-a for partial support. Their research was also supported
in parts by Russian Ministry of Education and Science, project No.
3.1386.2017.

\end{document}